\documentclass{emulateapj}
\slugcomment{The Astrophysical Journal, 697:1842-1860, 2009 June 1}
\shortauthors{Knobel et al.}
\begin{document}
\title{An optical group catalogue to z = 1 from the zCOSMOS 10k sample\altaffilmark{1}}

\author{C.~Knobel\altaffilmark{2},
S.~J.~Lilly\altaffilmark{2},
A.~Iovino\altaffilmark{7},
C.~Porciani\altaffilmark{2,17},
K.~Kova\v{c}\altaffilmark{2},
O.~Cucciati\altaffilmark{7},
A.~Finoguenov\altaffilmark{5},
M.~G.~Kitzbichler\altaffilmark{19}
C.~M.~Carollo\altaffilmark{2},
T.~Contini\altaffilmark{6},
J.-P.~Kneib\altaffilmark{8},
O.~Le~F\`{e}vre\altaffilmark{8},
V.~Mainieri\altaffilmark{4},
A.~Renzini\altaffilmark{13},
M.~Scodeggio\altaffilmark{9},
G.~Zamorani\altaffilmark{5},
S.~Bardelli\altaffilmark{5},
M.~Bolzonella\altaffilmark{5},
A.~Bongiorno\altaffilmark{3},
K.~Caputi\altaffilmark{2},
G.~Coppa\altaffilmark{5},
S.~de~la Torre\altaffilmark{8},
L.~de~Ravel\altaffilmark{8},
P.~Franzetti\altaffilmark{9},
B.~Garilli\altaffilmark{9},
P.~Kampczyk\altaffilmark{2},
F.~Lamareille\altaffilmark{6},
J.-F.~Le Borgne\altaffilmark{6},
V.~Le~Brun\altaffilmark{8},
C.~Maier\altaffilmark{2},
M.~Mignoli\altaffilmark{5},
R.~Pello\altaffilmark{6},
Y.~Peng\altaffilmark{2},
E.~Perez Montero\altaffilmark{6},
E.~Ricciardelli\altaffilmark{13},
J.~D.~Silverman\altaffilmark{2},
M.~Tanaka\altaffilmark{4},
L.~Tasca\altaffilmark{8},
L.~Tresse\altaffilmark{8},
D.~Vergani\altaffilmark{5},
E.~Zucca\altaffilmark{5},
U.~Abbas\altaffilmark{8,15},
D.~Bottini\altaffilmark{9},
A.~Cappi\altaffilmark{5},
P.~Cassata\altaffilmark{8},
A.~Cimatti\altaffilmark{10},
M.~Fumana\altaffilmark{9},
L.~Guzzo\altaffilmark{7},
A.~M.~Koekemoer\altaffilmark{20},
A.~Leauthaud\altaffilmark{18},
D.~Maccagni\altaffilmark{9},
C.~Marinoni\altaffilmark{11},
H.~J.~McCracken\altaffilmark{12},
P.~Memeo\altaffilmark{9},
B.~Meneux\altaffilmark{3,16},
P.~Oesch\altaffilmark{2},
L.~Pozzetti\altaffilmark{5},
R.~Scaramella\altaffilmark{14}
}

\altaffiltext{1}{European Southern Observatory (ESO), Large Program 175.A-0839}
\altaffiltext{2}{Institute for Astronomy, ETH Zurich, 8093 Zurich, Switzerland}
\altaffiltext{3}{Max-Planck-Institut f\"ur extraterrestrische Physik, D-84571 Garching, Germany}
\altaffiltext{4}{European Southern Observatory, Karl-Schwarzschild- Strasse 2, Garching, D-85748, Germany}
\altaffiltext{5}{INAF Osservatorio Astronomico di Bologna, via Ranzani 1, I-40127, Bologna, Italy}
\altaffiltext{6}{Laboratoire d'Astrophysique de Toulouse-Tarbes, Université de Toulouse, CNRS, 14 avenue Edouard Belin, F-31400 Toulouse, France} \altaffiltext{7}{INAF Osservatorio Astronomico di Brera, Milan, Italy}
\altaffiltext{8}{Laboratoire d'Astrophysique de Marseille, Marseille, France}
\altaffiltext{9}{INAF - IASF Milano, Milan, Italy}
\altaffiltext{10}{Dipartimento di Astronomia, Universit\'a di Bologna, via Ranzani 1, I-40127, Bologna, Italy}
\altaffiltext{11}{Centre de Physique Theorique, Marseille, Marseille, France}
\altaffiltext{12}{Institut d'Astrophysique de Paris, UMR 7095 CNRS, Universit\'e Pierre et Marie Curie, 98 bis Boulevard Arago, F-75014 Paris, France}
\altaffiltext{13}{Dipartimento di Astronomia, Universita di Padova, Padova, Italy}
\altaffiltext{14}{INAF, Osservatorio di Roma, Monteporzio Catone (RM), Italy}
\altaffiltext{15}{ELSA Marie Curie Postdoctoral Fellow, INAF - Osservatorio Astronomico di Torino, 10025 Pino Torinese, Italy}
\altaffiltext{16}{Universitats-Sternwarte, Scheinerstrasse 1, D-81679 Muenchen, Germany}
\altaffiltext{17}{Argelander-Institut f\"ur Astronomie, Auf dem H\"ugel 71, D-53121 Bonn, Germany}
\altaffiltext{18}{LBNL \& BCCP, University of California, Berkeley, CA, 94720}
\altaffiltext{19}{Max Planck Institute of Astrophysics, Karl-Schwarzschild Str. 1, PO Box 1317, D-85748 Garching, Germany}
\altaffiltext{20}{Space Telescope Science Institute, 3700 San Martin Drive, Baltimore, MD 21218}

\begin{abstract}

We present a galaxy group catalogue spanning the redshift range $0.1 \lesssim z \lesssim 1$ in the $\sim 1.7\ \rm{deg}^2$ COSMOS field, based on the first $\sim 10,000$ zCOSMOS spectra. The performance of both the Friends-of-Friends (FOF) and Voronoi-Delaunay-Method (VDM) approaches to group identification has been extensively explored and compared using realistic mock catalogues. We find that the performance improves substantially if groups are found by progressively optimizing the group-finding parameters for successively smaller groups, and that the highest fidelity catalogue, in terms of completeness and purity, is obtained by combining the independently created FOF and VDM catalogues. The final completeness and purity of this catalogue, both in terms of the groups and of individual members, compares favorably with recent results in the literature. The current group catalogue contains 102 groups with $N \geq 5$ spectroscopically confirmed members, with a further $\sim 700$ groups with $2 \leq N \leq 4$. Most of the groups can be assigned a velocity dispersion and a dark-matter mass derived from the mock catalogues, with quantifiable uncertainties. The fraction of zCOSMOS galaxies in groups is about $25 \%$ at low redshift and decreases toward $\sim 15 \%$ at $z \sim 0.8$. The zCOSMOS group catalogue is broadly consistent with that expected from the semi-analytic evolution model underlying the mock catalogues. Not least, we show that the number density of groups with a given intrinsic richness increases from redshift $z \sim 0.8$ to the present, consistent with the hierarchical growth of structure.

\end{abstract}

\keywords{catalogs --- galaxies: clusters: general --- galaxies: high-redshift --- methods: data analysis}

\section{Introduction} \label{sec:introduction}

Groups and clusters of galaxies are the most massive virialized structures in the Universe. They are important for several reasons. First, groups and clusters define the environment in which most galaxies actually reside and in which we may expect many important processes determining the evolution of galaxies (e.g., Voit 2005). Studying the properties of galaxies in groups at different redshifts is a direct probe of how the local environment affects the formation and evolution of galaxies with cosmic time. Second, characterization of galaxies in groups provides information about the galactic content of dark matter (DM) halos. This yields statistical quantities such as the halo occupation distribution (e.g., Collister \& Lahav 2005) or the conditional luminosity function (e.g., Yang et al. 2008) which themselves yield useful constraints on various physical processes that govern the formation and evolution of galaxies. Finally, the number density and clustering of groups strongly depend on cosmological parameters and thus are a potentially sensitive probe of the underlying cosmological model (e.g., Bahcall et al. 2003; Gladders et al. 2007; Rozo et al. 2009).

From an observational point of view, there are many ways to identify a group\footnote{In this paper, we will not distinguish between ``groups'' and ``clusters'', since from an optical/near-infrared point of view the difference between groups and clusters is rather a gradual, quantitative one, and not a qualitative one. So when we talk of ``groups'', we do not make any assumption about the mass or other properties oh these systems.}.  In the current $\Lambda$CDM-framework it is natural to associate groups with DM halos, and this is the definition adopted by most authors. Therefore, throughout this paper we refer to a ``group'' as a set of galaxies occupying the same DM halo\footnote{Throughout this paper, a DM halo is operationally defined as a friends-of-friends group of DM particles with a linking length of $b = 0.2$, since this is the definition adopted in the Millenium DM $N$-body Simulation (Springel et al. 2005) used for our analysis. So DM halos correspond to a mean overdensity of roughly 200. Alternative practical definitions or higher overdensities would then, in principle, correspond to different group catalogues.}.

There are many different observational techniques to identify groups in the local and distant Universe in use today. Groups can be detected in the optical/near-infrared (NIR) (e.g., Gal 2006), by diffuse X-ray emission (e.g., Pierre et al. 2006; Finoguenov et al. 2007), by the Sunyaev-Zel'dovich effect in the cosmic microwave background (e.g., Carlstrom et al. 2002; Voit 2005), by particular wide-angle tailed (WAT) galaxies (e.g., Blanton et al. 2003), and through cosmic shear due to weak gravitational lensing (e.g., Feroz et al. 2008). Each of these methods has its own advantages and problems (see e.g. Voit 2005; Johnston et al. 2007 \S~1), and the choice of a particular method might depend on the desired application.

If one aims to study the galaxy population in groups, searching for groups directly in large optical galaxy surveys is relatively straightforward and efficient. There are many different methods discussed in the literature to identify groups in an optical survey (for a review see e.g. Gerke et al. 2005 \S~4.1; Gal 2006). In essence, these aim to identify overdensities in redshift space, luminosity and/or color space, depending on the availability of redshift information and/or photometry. Whatever method is used, it should conform to the following general rules (see e.g. Gal 2006): First, it should be based on an objective, automated algorithm to minimize human biases. Second, the algorithm should impose minimal constraints on the physical properties of the clusters to avoid selection biases. The latter point is especially important if one aims to investigate the evolution of the galaxy population in groups. For instance, it has been shown that the addition of color information provides a powerful tool to find clusters in the Universe. There are methods such as the Cluster Red Sequence (CRS) method (Gladders \& Yee 2000) or the maxBCG algorithm (Hansen et al. 2005, Koester et al. 2007a) which are based on the fact that the most luminous galaxies in clusters inhabit a tight sequence in the color-magnitude diagram called the ``red sequence''. Using the red sequence information, these methods have proved to be very successful in finding clusters in the local (Koester et al. 2007b) and the distant Universe up to redshift $z \sim 1$ (Gladders \& Yee 2005). A further advantage of these methods is that no redshift information is needed. However, clearly the requirement of a substantial population of red sequence galaxies inhabiting the red sequence may impose a pre-selection that makes evolutionary studies more difficult.

The large number of accurate spectroscopic redshifts available for the large numbers of galaxies from the zCOSMOS redshift survey in the COSMOS
field (Scoville et al. 2007) enables us to use the most fundamental signature of groups -- overdensities in redshift space -- without recourse to additional color information. Nevertheless, even with precise spectroscopic redshifts, to identify groups in redshift space one has to deal with certain difficulties: Firstly, the peculiar velocities of galaxies in groups elongates groups in the redshift dimension (the ``fingers-of-god'' effect). This effectively decreases the galaxy density within groups in redshift space, and thus makes them harder to detect, and may cause group members to intermingle with other nearby field galaxies or even to merge into another nearby group. It is almost impossible to separate interlopers from real group galaxies if they appear within the group in redshift space. Second, in magnitude limited surveys such as ZCOSMOS, the mean density of galaxies decreases with redshift. So any algorithm based on the distance between neighbouring galaxies has to take into account the dependence of the mean galaxy separation with redshift. Third, the observational selection of galaxies (e.g. inhomogeneous sampling rate in the spectroscopic survey) frequently produce additional complications.

To cope with these difficulties, some forms of the traditional Friends-of-Friends (FOF) algorithm (Huchra \& Geller 1982) are still widely used (e.g. Eke et al. 2004; Berlind et al. 2006), although FOF has some well known shortcomings (e.g. Nolthenius \& White 1987; Frederic 1995). For instance, the FOF algorithm depends sensitively on the value of the linking length, and can merge neighbouring groups into single big groups, or fragment large groups into smaller pieces.

Until now, there have not been many spectroscopic redshift surveys searching for groups at high redshift. Carlberg et al. (2001) describe a group catalogue obtained from CNOC2 in the redshift range $0.1 \lesssim z \lesssim 0.5$. For the redshift range $z\gtrsim 0.5$ only the DEEP2 redshift survey (Davis et al. 2003), covering a total area of $\sim 3\ \rm{deg}^2$ and redshift range of $0.7 \lesssim z \lesssim 1.4$, has sufficient size and sampling rate to identify a large number of groups in redshift space. To achieve this, Gerke et al. (2005, 2007) have adapted the Voronoi-Delaunay-Method (VDM) of Marinoni et al. (2002), which is claimed to compensate for some of the shortcomings of the traditional FOF algorithm.

The aim of this paper is to create a group catalogue from the $\sim 10,000$ spectra in the zCOSMOS 10k sample (S. J. Lilly et al. 2009, in preparation) to enable the study of the group population over the redshift range $0.1 \lesssim z \lesssim 1$. We will compare the performance of both the FOF and VDM algorithms on the 10k sample, and try to optimize the group-finding methods by the introduction of a ``multi-run procedure''. In \S~2 we describe the 10k sample and corresponding realistic mock catalogues that were generated to test the group finding algorithms. \S~3 gives a detailed description of our adopted group-finding method, and discusses the performance of the two groupfinders. In \S~4 we present the 10k group catalogue, and describe how basic group properties are estimated. \S~5 compares the 10k group catalogue to the mocks and to 2dfGRS. Finally, \S~6 summarizes the paper. Where necessary, a concordance  cosmology with $H_0 = 73\ \rm{km\; s^{-1}\; \rm{Mpc}^{-1}}$, $\Omega_m = 0.25$, and $\Omega_\Lambda = 0.75$ is adopted. All magnitudes are quoted in the AB system.

\section{Data}

\subsection{zCOSMOS survey}

zCOSMOS is a spectroscopic redshift survey (Lilly et al. 2007, 2009 in preparation) covering the $\sim 1.7\ \rm{deg}^2$ COSMOS field (Scoville et al. 2007). The redshifts are measured with the VLT using the VIMOS spectrograph (Le F\`{e}vre et al. 2003). The zCOSMOS survey is split into two parts: The first part, ``zCOSMOS-bright'', is a pure magnitude selected survey with $15 \leq \rm{I_{AB}} \leq 22.5$, $\rm{I_{AB}}$ the F814W HST/ACS band (Koekemoer et al. 2007). This magnitude limit will yield a survey of approximately 20,000 galaxies in the redshift range $0.1 \lesssim z \lesssim 1.2$. Repeated observations of some zCOSMOS galaxies have shown that the redshift error is approximately Gaussian distributed with a standard deviation of $\sigma_v \simeq 100\ \rm{km\ s^{-1}}$. The second part of zCOSMOS, ``zCOSMOS-deep'', aims at observing about 10,000 galaxies in the redshift range $1.5 \lesssim z \lesssim 3.0$ selected through a well-defined color criteria. 

To date, about a half of zCOSMOS-bright has been completed yielding about $10,500$ spectra (S. J. Lilly et al. 2009, in preparation). Among these redshifts about $15 \%$ are classified as unreliable. For the group catalogue, we have accepted all objects with the confidence classes 4 and 3, 9.5, 9.3, 2.5, 9.4, 2.4, 1.5, and 1.4 (see S. J. Lilly et al. 2009, in preparation). The redshifts with these confidence classes constitute $86 \%$ of the whole 10k sample and have a spectroscopic confirmation rate of $98.6 \%$ as found by duplicate observations. After removing the stars ($\sim 5 \%$), we finally end up with a sample of 8417 galaxies with usable redshifts (``10k sample'').

At the current stage of the survey, the spatial spectroscopic sampling rate of galaxies across the COSMOS field is very inhomogeneous, and there are clearly some linear features such as stripes visible (see Figure~5 of S. J. Lilly et al. 2009, in preparation). Since this will affect the number of detectable groups in this sample in a non-trivial way, we have created mock catalogues that have the same kind of inhomogeneous coverage. To create the group catalogue and generate the statistics describing the fidelity of the catalogue, the groupfinders were applied to the whole field spanning the range $149.47^\circ \lesssim \alpha \lesssim 150.77^\circ$ and $1.62^\circ \lesssim \delta \lesssim 2.83^\circ$. However, for some applications discussed below we restrict ourselves to the ``central region'' of the COSMOS field defined by $\alpha = 150\pm 0.4^\circ$ and $\delta = 2.15\pm 0.4^\circ$, since this region is relatively complete compared to the total field. Only about $25 \%$ of the area has a completeness lower than $30 \%$ while for the whole field this area constitutes more than $50 \%$.

\begin{figure}
	\epsscale{1}\plotone{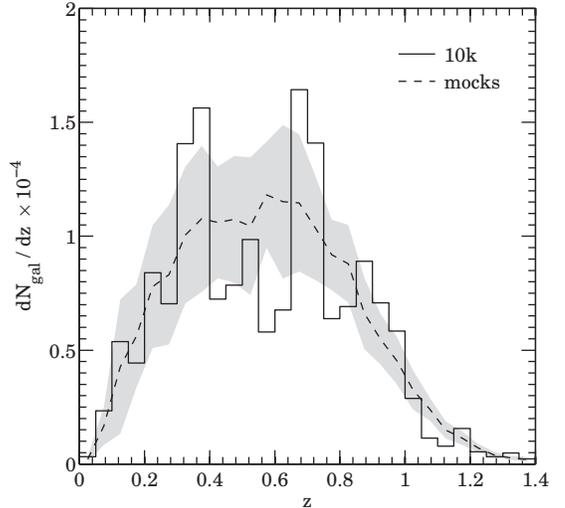}
		\figcaption{Number of galaxies per redshift $dN_{\rm gal}/dz$. The histogram shows the $dN_{\rm gal}/dz$ of the 10k sample used in this paper. Two large over-densities at $z \sim 0.3$ and $z \sim 0.7$ are clearly visible. The dashed line shows the mean $dN_{\rm gal}/dz$ of the 24 mocks and the shaded area their scatter. As noted below, the magnitude limit in the mocks have been adapted such that the mean $dN_{\rm gal}/dz$ of the mocks matches the smoothed $dN_{\rm gal}/dz$ of HST/ACS COSMOS catalogue. The shaded area shows that, although COSMOS covers an unprecedentedly large area for a survey of this depth, cosmic variance is still an important issue.\label{fig:zCOSMOS_number}}
\end{figure}

The number of galaxies per unit redshift $dN_{\rm gal}/dz$ is shown in Figure~\ref{fig:zCOSMOS_number}. There are two striking density peaks at redshifts $z \sim 0.3$ and $\sim 0.7$.

\subsection{zCOSMOS 10k mocks} \label{sec:mock}

The mocks we use to calibrate and test our groupfinders are adapted from the COSMOS mock lightcones (Kitzbichler \& White 2007). These light cones are based on the Millenium DM $N$-body simulation (Springel et al. 2005) which was run with the cosmological parameters $\Omega_m = 0.25$, $\Omega_\Lambda = 0.75$, $\Omega_b = 0.045$, $h = 0.73$, $n = 1$, and $\sigma_8 = 0.9$. The semi-analytic recipes for populating the volume with galaxies in the lightcones is that of Croton et al. (2006) as updated by De Lucia \& Blaizot (2006). There are 24 independent mocks, each covering an area of $1.4\ \rm{deg} \times 1.4\ \rm{deg}$ with an apparent magnitude limit of $r \leq 26$ and galaxies in the redshift range $z \lesssim 7$.

These lightcones were adjusted to resemble the real 10k sample as much as possible. First, a magnitude cut of $15 \leq i \leq 22.5$ was applied. However, the mean number of galaxies in the resulting mocks was about $5-10 \%$ higher than in the zCOSMOS target catalogue (i.e. a $1-2\sigma$ effect). To make the mocks more closely resemble the real data, we adjusted the magnitude cut in a redshift dependent way so that the mean number of galaxies per unit redshift $\bar{N_{\rm gal}}(z)/dz$ in the mocks was equal to the smoothed $N_{\rm gal}(z)/dz$ of the zCOSMOS input target catalogue (see Figure~\ref{fig:zCOSMOS_number}). Then, the spatial sampling completeness and the redshift success rate were simulated by removing galaxies from the mocks according to the probability that a galaxy with a certain position and redshift would have been observed in the 10k sample. It should be noted that zCOSMOS is a slit-based survey.  However, the bias against close neighbours --- already small because of the multiple passes (upto 8 in the central region) across the field --- is further mitigated by galaxies appearing serendipitously in slits targeted at other galaxies (see P. Kampczyck et al. 2009, in preparation). The small variation in sampling rate on these small scales, which is anyway well below the mean intergalactic separation in 3-d space, has been ignored in constructing the mocks. To further enhance the conformity with the 10k sample, the redshift of each galaxy was perturbed by an amount drawn from a Gaussian distribution with standard deviation $\sigma_z = 100(1+z)/c\ \rm{km\ s^{-1}}$.  

\subsection{Detectability of groups}\label{sec:detectability}

Since, according to our definition, a group is the set of galaxies occupying the same DM halo, we can only hope to detect those groups which host at least two galaxies in the 10k sample. The collection of all these ``detectable groups'' constitutes the ideal (or ``real'') group catalogue.  This is the best catalogue that can be produced with the 10k sample, and this is the catalogue we aim to reconstruct with our groupfinder. Any DM Halo hosting only a single zCOSMOS 10k galaxy is not detectable and the corresponding galaxies will be termed ``field galaxies''.  For this reason, even the ideal group catalogue that is detectable with the 10k sample will not be a complete rendition of the true underlying group population in the COSMOS volume.  Nevertheless, whenever we discuss the statistical properties of a group catalogue, such as completeness or purity (see \S~\ref{sec:statistics}), these will be measured relative to this ``ideal'' group catalogue, rather than the underlying population.

\begin{figure}
		\epsscale{1}\plotone{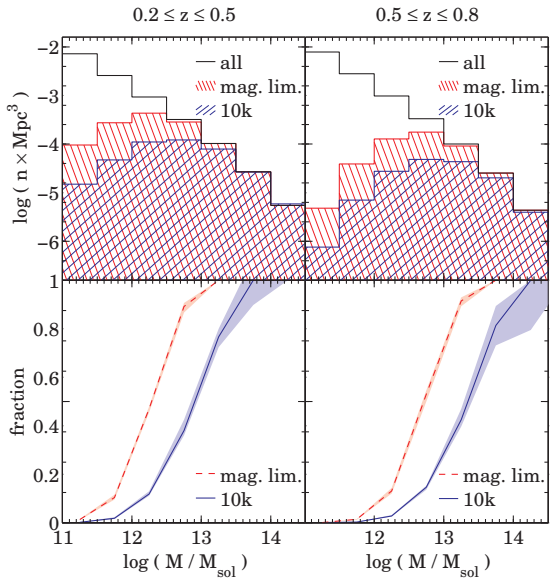}
		\figcaption{Fraction of detectable groups in the ``ideal'' mocks as a function of DM halo mass. The left panel
shows the redshift range $0.2 \leq z \leq 0.5$, and the right panel shows $0.5 \leq z \leq 0.8$. The upper panels show the number density of halos in the 10k mocks (blue), in a purely 22.5 magnitude limited sample (red), and in total (black), and the lower panels show the fraction of halos in the 10k sample (solid line) and in the magnitude limited sample (dashed line) with respect to the total number of halos at a given mass. The shaded regions show the upper and lower quartiles of the fractions among the 24 mocks. For both redshift ranges, the 10k sample was restricted to the central region of the survey.\label{fig:halo_incompleteness}}
\end{figure}

In a flux limited survey such as zCOSMOS, the population of galaxies that is observed changes with redshift, and the same will also therefore be true of the groups. For instance, for a group to be detectable at high redshift, it has to host at least two rather bright galaxies. Figure \ref{fig:halo_incompleteness} shows the fraction of detectable groups in the mocks (i.e. in the ``ideal'' catalogue in the previous paragraph) as a function of the halo mass in the two redshift bins $0.2 \leq z \leq 0.5$ and $0.5 \leq z \leq 0.8$.
While in the lower redshift bin the sample should be complete down to $\sim 5\times 10^{13} M_{\odot}$, this limit increases in the higher redshift bin to $\sim 2\times 10^{14} M_{\odot}$. However, in both bins the bulk of the detectable halos are in the mass range $10^{12} M_{\odot} \lesssim M \lesssim 5\times 10^{13} M_{\odot}$.

\section{group-finding method} \label{sec:group-finding}

In this section the different group-finding methods and the statistical properties of the resulting group catalogues are discussed. We have applied both the FOF and VDM algorithms to our sample. In this way we are able to compare the resulting group catalogues obtained by the different methods and to investigate the robustness of the results.

\subsection{The FOF and VDM algorithms} \label{groupfinder}

\subsubsection{FOF}

The FOF algorithm is adopted from Eke et al. (2004). It has three free parameters: the linking length $b$, the maximum perpendicular linking length in physical coordinates $L_{\rm max}$, and the ratio between the linking length along and perpendicular to the line of sight $R$. The exact meaning of these parameters becomes clear by regarding the linking criteria: Consider two galaxies $i$ and $j$ with comoving distances $d_i$ and $d_j$ respectively. These two galaxies are assigned to the same group if their angular separation $\theta_{ij}$ satisfies
\begin{eqnarray}
\theta_{ij} \leq \frac{1}{2} \left(\frac{l_{\perp,i}}{d_i}+\frac{l_{\perp,j}}{d_j}\right)
\end{eqnarray}
and, simultaneously, the difference between their distances satisfies
\begin{eqnarray}
\left|d_i-d_j\right| \leq \frac{l_{\parallel,i}+l_{\parallel,j}}{2}.
\end{eqnarray}
$l_{\perp}$ and $l_{\parallel}$ are the comoving linking lengths perpendicular and parallel to the line of sight defined by
\begin{eqnarray}
l_{\perp} &=& \min \left[L_{\rm max}(1+z),\frac{b}{\bar{n}^{1/3}}\right]\\
l_{\parallel} &=& R\; l_{\perp},
\end{eqnarray}
where $\bar{n}$ is the mean density of galaxies. Since the sample of galaxies is magnitude limited, the mean density of galaxies decreases with redshift leading to a steady increase of the mean inter-galaxy separation with redshift. Eke et al. (2004) argued that scaling both $l_\perp$ and $l_\parallel$ with $n^{-1/3}$ will compensate for the magnitude limit and lead to groups of similar shape and overdensity throughout the survey. The free parameter $L_{\rm max}$ has been introduced to avoid unphysically large values for $l_\perp$ at high redshifts where the galaxy distribution is sampled very sparsely. Since $L_{\rm max}$ is measured in physical coordinates, $L_{\rm max}(1+z)$ is the maximal comoving linking length perpendicular to the line of sight. Finally, the free parameter $R$ allows $l_\parallel$ to be larger than $l_\perp$ taking into account the elongation of groups along the line of sight due to the fingers-of-god effect.

\subsubsection{VDM}\label{seq:VDM}

The VDM algorithm was adopted from Gerke et al. (2005) which was itself based on the method developed by Marinoni et al. (2002). This algorithm is more complicated than the FOF and has six free parameters instead of three. Basically one needs a full Voronoi-Delaunay tesselation\footnote{For a given set of sites in space, the ``Voronoi-cell'' of a certain site consists of all points closer to this site than to any other site. Furthermore, two sites whose Voronoi-cells share a common interface are called ``Delaunay-neighbours''. The ``Voronoi-Delaunay tesselation'' for a given set of sites is the complete set of all its Voronoi-cells and Delaunay-neighbours. For more formal definitions and basic properties of Voronoi-Delaunay tesselations we refer to basic textbooks of geometry.} of the input galaxy sample and the volumes of each Voronoi-cell. The Voronoi-Delaunay-tesselation was computed using Qhull\footnote{http://www.qhull.org} (Barber, Dobkin, \& Huhdanapaa 1996) and the volumes of the Voronoi-cells using the algorithm of Mirtich (1996).

The VDM algorithm can be divided into 3 phases: In Phase I, the galaxies are ordered in ascending order of Voronoi-volumes. Then, the first galaxy in this sorted list is taken as a ``seed galaxy'' and a cylinder of radius $R_{\rm I}$ and length $2L_{\rm I}$ using comoving coordinates is placed around it such that the axis of the cylinder is directed along the line of sight. If there is no other galaxy inside this cylinder, the ``seed galaxy'' is regarded as a field galaxy and one proceeds to the next galaxy in the list. If, however, there are other galaxies within the cylinder, Phase II starts. In this phase, a second cylinder with radius $R_{\rm II}$ and length $2L_{\rm II}$ is defined and all galaxies inside this second cylinder directly connected to the seed galaxy or to its immediate Delaunay-neighbours by means of the Delaunay-mesh are assigned to the same group. The number of galaxies inside the second cylinder $N_{\rm II}$ is taken as an estimate of the central richness of the group. In Phase III, a third cylinder with radius 
\begin{eqnarray}
R_{\rm III} &=& r(\tilde N_{\rm II})^{1/3}\\
L_{\rm III} &=& l(\tilde N_{\rm II})^{1/3} f(z)
\end{eqnarray}
is defined, whereas $r$ and $l$ are two free parameters, $\tilde N_{\rm II}$ is the central richness corrected for the redshift dependent mean density $\bar{n}(z)$, and $f(z)$ is a function introduced to take into account that for a fixed velocity dispersion the length of the fingers of god in redshift space is a function of redshift. $\tilde N_{\rm II}$ and $f(z)$ are given by
\begin{eqnarray}
\tilde N_{\rm II} = \frac{\bar{n}(z_{\rm ref})}{\bar{n}(z)} N_{\rm II}
\end{eqnarray}
and
\begin{eqnarray}
f(z) = \frac{s(z)}{s(z_{\rm ref})},\ \ \ s(z) = \frac{1+z}{\sqrt{\Omega_m(1+z)^3+\Omega_\Lambda}}
\end{eqnarray}
respectively, where $z_{\rm ref}$ is an arbitrary reference redshift chosen to be 0.5. In this third phase, again all galaxies within the third cylinder are assigned to the current group. After fixing $z_{\rm ref}$, 6 free parameters $R_{\rm I}$, $L_{\rm I}$, $R_{\rm II}$, $L_{\rm II}$, $r$, and $l$ remain. The reader is referred to Table~\ref{tab:optimal-parameter-sets} for typical parameter-sets for the two group-finders.

It can be seen that both group-finding algorithms are somewhat arbitrary and neither is directly inked to the physical basis of a group, namely virialized motion within a common potential well. While it seems that the VDM algorithm is at least partly motivated by certain scaling relations for groups (Gerke et al. 2005), this is at the expense of simplicity which is clearly the mark of the FOF algorithm. 

\subsection{Basic statistical quantities}
\label{sec:statistics}

In order to assess the performance of a groupfinder, realistic mock catalogues containing full information about the underlying DM halos and their properties are needed. In this section, we introduce some useful statistics to characterize the overall fidelity of the resulting group catalogues.

\begin{figure*}
		\epsscale{1}\plotone{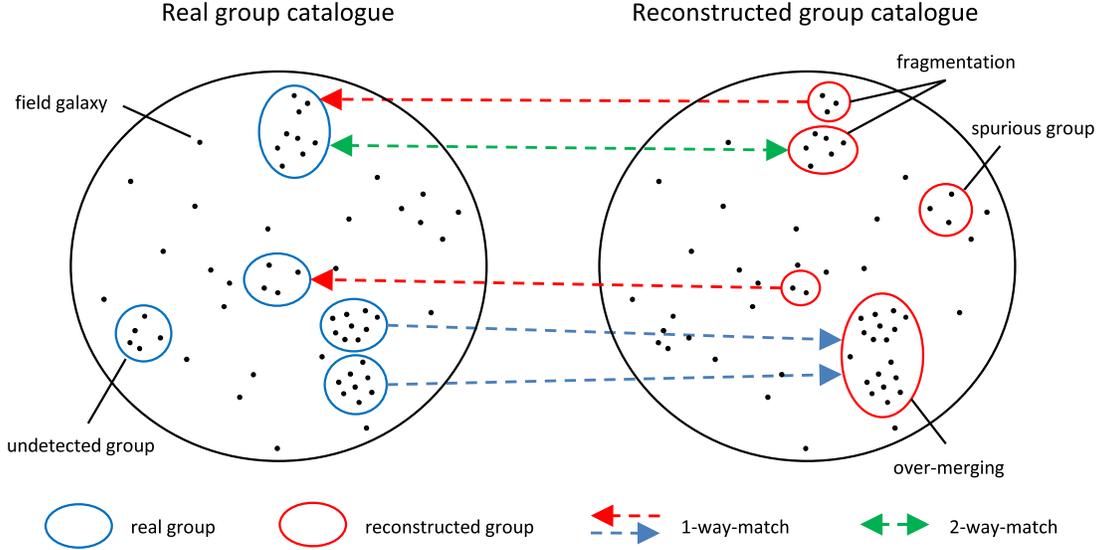}
		\figcaption{Schematic illustration of comparing a reconstructed group catalogue to a real group catalogue as obtained from DM simulation. The left big circle constitutes the real group catalogue and the right big circle the reconstructed group catalogue. Each point displays a galaxy and the encircled points inside the big circles constitute groups. A group in the real (reconstructed) catalogue may be associated to a group in the reconstructed (real) catalogue (see the text for details). Such an association is indicated by an arrow pointing from the real (reconstructed) group to the reconstructed (real) group. If there is an arrow pointing from one group to another and also an arrow pointing backwards, such an association is termed a ``two-way-match''. Otherwise it is just a ``one-way-match''. If more than one reconstructed group points to the same real group this is called ``fragmentation'', if there is more than one real group associated to the same reconstructed group, this is called ``over-merging''.\label{fig:arrows}}
\end{figure*}

The fidelity of the group catalogue can be assessed through comparing the ``reconstructed'' groups, obtained by running the groupfinder on the mock catalogues, to the ``real'' group catalogue described above - i.e. the set of all DM haloes in the mocks that contain, after the 10k selection criteria have been applied, at least two galaxies. The comparison is therefore of two identical point sets, the galaxies in the mocks, whose points are grouped together in possibly different ways. This is schematically illustrated in Figure \ref{fig:arrows}.
\begin{deluxetable*}{cccccccccccc}
\tablewidth{0.7\textwidth}
\tablecaption{Optimal multi-run parameter-sets for FOF and VDM}
\tablehead{
\colhead{} & \colhead{} & \multicolumn{3}{c}{FOF} & \colhead{} & \multicolumn{6}{c}{VDM} \\ 
\cline{3-5} \cline{7-12} \\ 
	\colhead{step} &
	\colhead{} &
  \colhead{$b$} &
  \colhead{$L_{\rm max}$} &
  \colhead{$R$} &
  \colhead{} &
  \colhead{$R_I$} &
  \colhead{$L_I$} &
  \colhead{$R_{II}$} &
  \colhead{$L_{II}$} &
  \colhead{$r$} &
  \colhead{$l$} \\
  \colhead{} &
	\colhead{} &
  \colhead{} &
  \colhead{(Mpc)\tablenotemark{a}} &
  \colhead{} &
  \colhead{} &
  \colhead{(Mpc)\tablenotemark{b}} &
  \colhead{(Mpc)\tablenotemark{b}} &
  \colhead{(Mpc)\tablenotemark{b}} &
  \colhead{(Mpc)\tablenotemark{b}} &
  \colhead{(Mpc)\tablenotemark{b}} &
  \colhead{(Mpc)\tablenotemark{b}}
  }
\startdata

1 && 0.11 & 0.45 & 13 &&  0.7 & 10 & 1.0 & 8  & 0.6 & 10 \\
2 && 0.11 & 0.45 & 13 &&  0.7 & 10 & 0.4 & 12 & 0.6 & 7  \\
3 && 0.08 & 0.4  & 12 &&  0.5 & 6  & 0.2 & 12 & 0.5 & 7  \\
4 && 0.19 & 0.4  & 11 &&  0.4 & 10 & 0.4 & 12 & 0.5 & 4  \\
5 && 0.07 & 0.3  & 18 &&  0.4 & 8  & 0.6 & 8  & 0.5 & 7  

\enddata

\tablenotetext{a}{physical coordinates}
\tablenotetext{b}{comoving coordinates}
\tablecomments{The definitions of these parameters are given in \S~\ref{groupfinder}.}

\label{tab:optimal-parameter-sets}
\end{deluxetable*}

We follow here the definitions and notations of Gerke et al. (2005). The two big circles constitute two group catalogues. Each point corresponds to a galaxy of the input galaxy sample and the encircled galaxies belong to the same group. In the left-hand catalogue are the ``real groups'' as given by the DM halos in the simulation, while in the right-hand catalogue are the ``reconstructed groups'' as identified by our groupfinder. Some sort of measure is needed of how many reconstructed groups can be identified with real groups and how many real groups are recovered by our groupfinder. Following Gerke et al. (2005), we define the following terms:

\begin{description}
\item[]\emph{Association:} A group $i$ is associated to another group $j$ if group $j$ contains more than the fraction $f$ of the members of group $i$. For this association to be unique, it must hold $f \geq 0.5$. Throughout this paper, we set $f = 0.5$ as did Gerke et al. (2005).
\item[]\emph{One-way-match:} If group $i$ is associated to group $j$, but group $j$ is not associated to group $i$ (illustrated by an arrow from group $i$ to group $j$).
\item[]\emph{Two-way-match:} If group $i$ is associated to group $j$ and vice versa (illustrated by a double-arrow).
\end{description}
While each group can only have a single, unique associated group (i.e. an arrow pointing away), it might well happen that a certain group is the associated group for many other groups (i.e. many arrows pointing toward it). We therefore have the following terminology:
\begin{description}
\item[]\emph{Over-merging:} If more than one real group is associated to the same reconstructed group.
\item[]\emph{Fragmentation:} If more than one reconstructed group is associated to the same real group.
\item[]\emph{Spurious group:} A reconstructed group which has no associated real group.
\item[]\emph{Undetected group:} A real group which has no associated reconstructed group.
\item[]\emph{Group galaxy:} A galaxy which belongs to a group.
\item[]\emph{Field galaxy:} A galaxy not associated to any group.
\end{description}

With this terminology, the following statistical measures can be defined that together describe the overall fidelity of the reconstructed group catalogue and thus its potential usefulness for quantitative analysis.  Let $N^{\rm real}_{\rm gr}(N_{\rm real})$ denote the number of real groups with $N_{\rm real}$ members, and $N^{\rm rec}_{\rm gr}(N_{\rm rec})$ the number of reconstructed groups with $N_{\rm rec}$ members. Then by
\begin{eqnarray}
\mathcal{A}[N_{\rm gr}^{\rm real}(N_{\rm real}) \rightarrow N_{\rm gr}^{\rm rec}(N_{\rm rec})] 
\end{eqnarray}
we denote the number of associations of real groups with $N_{\rm real}$ members to reconstructed groups with $N_{\rm rec}$ members. In the same way,
\begin{eqnarray}
\mathcal{A}[N_{\rm gr}^{\rm rec}(N_{\rm rec}) \rightarrow N_{\rm gr}^{\rm real}(N_{\rm real})]
\end{eqnarray}
denotes the number of associations of reconstructed groups with $N_{\rm rec}$ members to real groups with $N_{\rm real}$ members. The analogue notations for the numbers of two-way-associations are
\begin{eqnarray}
\mathcal{A}[N_{\rm gr}^{\rm real}(N_{\rm real}) &\leftrightarrow& N_{\rm gr}^{\rm rec}(N_{\rm rec})]\\
\mathcal{A}[N_{\rm gr}^{\rm rec}(N_{\rm rec}) &\leftrightarrow& N_{\rm gr}^{\rm real}(N_{\rm real})].
\end{eqnarray}
Note that the last two expressions are equivalent to each other. Then, with these notations we can formally introduce the ``one-way completeness'' $c_1(N)$ and the ``two-way completeness'' $c_2(N)$ by
\begin{eqnarray}
c_1(N) &=& \frac{\mathcal{A}[N_{\rm gr}^{\rm real}(\geq \!\! N) \rightarrow N_{\rm gr}^{\rm rec}(\geq \! 2)]}{N_{\rm gr}^{\rm real}(\geq \!\! N)}\\
c_2(N) &=& \frac{\mathcal{A}[N_{\rm gr}^{\rm real}(\geq \!\! N) \leftrightarrow N_{\rm gr}^{\rm rec}(\geq \! 2)]}{N_{\rm gr}^{\rm real}(\geq \!\! N)}.
\end{eqnarray}
Analogously, we define the ``one-way purity'' $p_1(N)$ and ``two-way purity'' $p_2(N)$ as
\begin{eqnarray}
p_1(N) &=& \frac{\mathcal{A}[N_{\rm gr}^{\rm rec}(\geq \!\! N) \rightarrow N_{\rm gr}^{\rm real}(\geq \! 2)]}{N_{\rm gr}^{\rm rec}(\geq \!\! N)}\\
p_2(N) &=& \frac{\mathcal{A}[N_{\rm gr}^{\rm rec}(\geq \!\! N) \leftrightarrow N_{\rm gr}^{\rm real}(\geq \! 2)]}{N_{\rm gr}^{\rm rec}(\geq \!\! N)}.
\end{eqnarray}
The one-way ``completeness'' $c_1(N)$ is a measure of the fraction of real groups with $N$ or more members that are successfully recovered in the reconstructed group catalogue, and the one-way ``purity'' $p_1(N)$ is a measure of the fraction of reconstructed groups with $N$ or more members that belong to real groups. The higher $c_1(N)$ the smaller the fraction of undetected groups ($1-c_1(N)$), and the higher $p_1(N)$ the smaller the fraction of spurious groups ($1-p_1(N)$). On the other hand, the smaller the ratios $c_2(N) / c_1(N)$ or $p_2(N) / p_1(N)$ the more over-merging or fragmentation, respectively, is present. By definition the four quantities $c_1(N)$, $c_2(N)$, $p_1(N)$, and $p_2(N)$ all take only values between 0 and 1.

While Gerke et al. (2005) have introduced these four quantities $c_1$, $c_2$, $p_1$, and $p_2$ globally for a group catalogue including all groups, we have defined them to be functions of the number of members $N$ (``richness''). It will become clear below that investigating these statistics as a function of $N$ is very useful for improving the performance of a group catalogue. Note that the argument $N$ always means ``for groups with $N$ or more members'' as is clear from their definitions, so for $N=2$ the two definitions are identical. Throughout this paper we will always consider the set of groups down to a given richness-class $N$. So this convention eases the notation. It would, however, be straightforward to define the analogue quantities in a non-cumulative way.

While $c_1(N)$, $p_1(N)$ etc. are statistical quantities on a group-to-group basis, statistical quantities on a galaxy-to-group basis may be useful as well. Therefore, following Gerke et al. (2005), we define the ``galaxy success rate'' $S_{\rm gal}(N)$ and the ``interloper fraction'' $f_{\rm I}(N)$ as
\begin{eqnarray}
S_{\rm gal}(N) &=& \frac{[S^{\rm gal}_{\rm real}(\geq \!\! N) \cap S^{\rm gal}_{\rm rec}(\geq \! 2)]}{[S^{\rm gal}_{\rm real}(\geq \!\! N)]} \\
f_{\rm I}(N) &=& \frac{[S^{\rm gal}_{\rm rec}(\geq \!\! N) \cap S^{\rm gal}_{\rm field}]}{[S^{\rm gal}_{\rm rec}(\geq \!\! N)]},
\end{eqnarray}
where $S^{\rm gal}_{\rm real}(N)$ is the set of galaxies associated to real groups of $N$ members, $S^{\rm gal}_{\rm rec}(N)$ the set of galaxies associated to reconstructed groups of $N$ members, and $S^{\rm gal}_{\rm field}$ the set of real field galaxies. The square brackets $[.]$ here denote the number of elements in a set and the $\cap$ is the usual intersection from set theory. Thus galaxy success rate $S_{\rm gal}(N)$ is just the fraction of galaxies belonging to real groups of richness $\geq \!\! N$ that have ended up in any reconstructed group, and the interloper fraction $f_{\rm I}(N)$ is the fraction of galaxies belonging to reconstructed groups of richness $\geq \!\! N$ that are field galaxies (``interlopers''). Like $c_1(N)$, $p_1(N)$, etc., $S_{\rm gal}(N)$ and $f_{\rm I}(N)$ will also take values between 0 and 1.   

\begin{figure*}
		\epsscale{1}\plotone{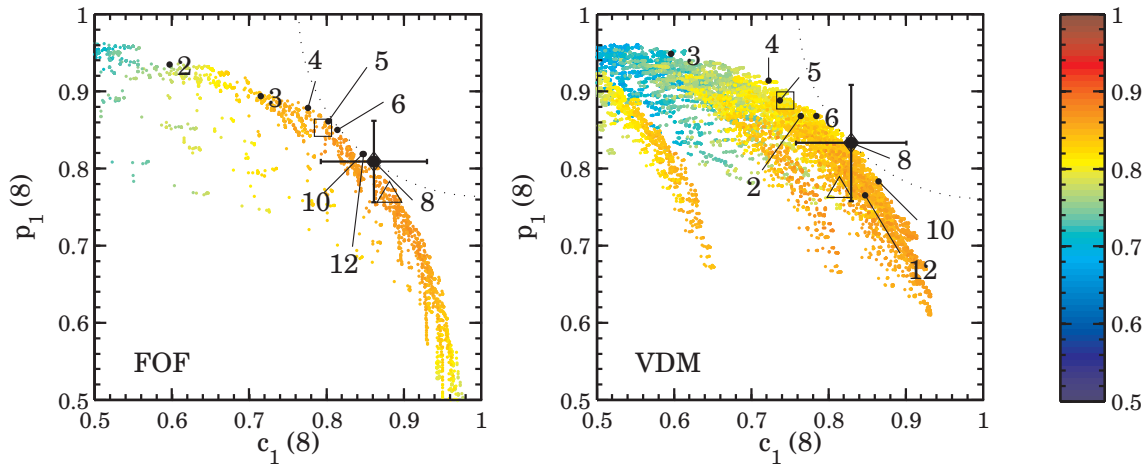}
		\figcaption{Distributions of parameter-sets in the $c_1(8)$-$p_1(8)$-plane for a wide range of group-finding parameters. In the left panel are the parameter-sets for FOF and in the right panel those for VDM. Each parameter-set is positioned at the average value for the 24 separate mock catalogues. The parameter-sets are color coded by the goodness parameter $g_2(8)$ indicating the degree of over-merging or fragmentation. The dotted line is the largest circle around the upper right corner being empty of points, i.e. the radius of this circle is equal to the smallest $g_1(8)$ value. The best $g_1(8)$ parameter-set is marked by a diamond and the error bars exhibit the scatter among the 24 mocks for this particular parameter-set. The labeled black points show the sites where the best $g_1(N)$ sets for different $N$ reside on this plane, $N$ being denoted by the label of the points. Although these best sets inhabit, in general, very different places, they converge for $N\geq 8$, at least for FOF. The position of the best $g_2(8)$-set is marked by a triangle and the one of the best $g_3(8)$ by a square.\label{fig:parameter_scatter}}
\end{figure*}

It is well known (e.g. Frederic 1995, Gerke et al. 2005) that a perfect reconstructed group catalogue is impossible to achieve and furthermore, that completeness and purity tend to be mutually exclusive. As would be expected, the higher the completeness, the lower the purity, and vice versa (see Figure \ref{fig:parameter_scatter}).
There is also a similar dichotomy between over-merging and fragmentation. Therefore, we introduce additional measures of ``goodness'' which combine the statistics such as completeness and purity in a way that maximizing (or minimizing) them yields a sort of ``optimal'' group catalogue. We formally define as (omitting the dependence of $N$ for the sake of clarity):
\begin{eqnarray}
g_1 &=& \sqrt{(1-c_1)^2 + (1-p_1)^2} \label{eq:g1}\\
g_2 &=& \frac{c_2}{c_1}\;\frac{p_2}{p_1}\\
g_3 &=& \sqrt{(1-S_{\rm gal})^2 + f_{\rm I}^2}.
\end{eqnarray}
The meaning of these quantities is as follows: Since a perfect group catalogue features $(c_1,p_1) = (1,1)$, i.e. entirely complete and absolutely pure, the reconstructed group catalogue should come as close as possible to this point in the $c_1$-$p_1$-plane. So $g_1$ gives the distance to this optimal point in the $c_1$-$p_1$-plane and thus is a measure of the balance of completeness and purity. Then, a good group catalogue should exhibit $c_1 \simeq c_2$ and $p_1 \simeq p_2$ meaning that essentially no over-merging and fragmentation is present in the catalogue. Hence, $g_2$ measures the balance between over-merging and fragmentation and should also approach 1. Finally, $g_3$ is similar to $g_1$ but is on a galaxy-to-group basis instead of a group-to-group basis. As is clear from their definitions, these measures of goodness again take only values between 0 and 1. It is clear that $g_1$ and $g_3$ should be minimized, while $g_2$ should be maximized.

\subsection{Optimization strategy} \label{sec:optimization_strategy}

Since there exists no single perfect reconstructed group catalogue, one has to optimize the group-finding parameters, in principle, in a way that the resulting group catalogue serves as well as possible the intended scientific purpose. However, as we will see, there seems to be a rather natural way to construct a group catalogue which is useful for many different purposes. The only way to find such optimal parameters of a groupfinder is to run it on the mocks for different parameter-sets, and to compare the resulting group catalogues by means of the statistics introduced in the previous section.

The completeness $c_1(8)$ and purity $p_1(8)$ of the reconstructed group catalogues, after running FOF and VDM over a large parameter space, are shown in Figure \ref{fig:parameter_scatter}. It is obvious that the points do not extend arbitrarily close to the right upper corner (i.e. the perfect group catalogue). The parameters $c_1(8)$ and $p_1(8)$ are in some sense anti-correlated. In fact, the cloud of points seem to feature a boundary toward high completeness and purity beyond which there is a region totally free of points. It is notable, how similar this boundary is for FOF and VDM approaches --- clearly neither is markedly superior to the other. The same holds for the $g_2(8)$-goodness, color coded in the figure, along this boundary region. These similarities between FOF and VDM are observed for all richness classes $N$. This indicates that this boundary is probably the limit of what can be achieved with a zCOSMOS-10k-like sample and does not depend on the choice of algorithm. This also suggests that the choice of a particular groupfinder such as FOF or VDM is less important than sometimes argued, although, as we will see, the properties of group catalogues obtained using the two groupfinders are not absolutely identical.

VDM, much more than FOF, also exhibits some scatter in the range given by $0.5 < c_1(8) < 0.85$ and $p_1(8) > 0.65$. The existence of such parameter-sets is a natural side-effect of the relatively large number of free parameters of the VDM groupfinder resulting in many parameter combinations with obviously suboptimal properties in terms of $c_1(8)$ and $p_1(8)$. The extent of this scatter, of course, also depends strongly on the explored range of values in the parameter space. Since we are interested in parameter-sets yielding simultaneously high completeness and high purity, we will only focus on the boundary mentioned above.

The challenge is to find the best group catalogues among those plotted in Figure \ref{fig:parameter_scatter}, making the best compromise between $c_1$ and $p_1$. A natural choice is the point that lies closest to $(c_1,p_1) = (1,1)$ indicated by the diamond. According to equation (\ref{eq:g1}), this is the point where $g_1$ is minimal. We will refer to this parameter-set as the ``best $g_1$-set''. It defines a circle around the upper right corner (dotted line) that is entirely empty of points.

In addition to minimizing $g_1$, one would prefer, of course, to simultaneously maximize $g_2$ and minimize $g_3$. In general, the best parameter-sets for these three goodnesses will not coincide. Rather it turns out that the best $g_2$-set lies usually at slightly higher completeness relative to the best $g_1$-set (see triangles in Figure \ref{fig:parameter_scatter}), while the best $g_3$-set lies usually at slightly lower completeness (see squares in Figure \ref{fig:parameter_scatter}). However, as is clear from Figure \ref{fig:parameter_scatter}, the gradient of $g_2$ is rather shallow around the best $g_1$-set and nearly maximal, so that the precise site of the optimal $g_2$-set is not that important. The same holds for the gradient of $g_3$. Finally, it seems that the best $g_1$-set is a good choice.

\subsubsection{Multi-run procedure} \label{seq:multi_run_procedure}

\begin{figure}
		\epsscale{1}\plotone{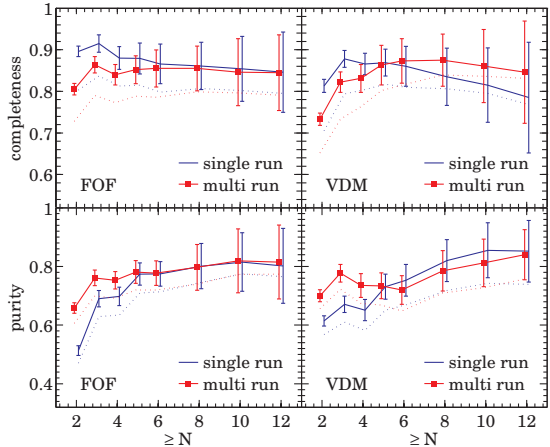}
		\figcaption{Comparison of the completeness and purity obtained from a single run and from the multi-run-procedure. The left two panels show the statistics for FOF and the right two panels for VDM. In each panel, the blue color corresponds to the single run and the red color to the multi-run-procedure. In the upper two panels, the solid lines display the one-way completeness $c_1$, and in the lower two panels they show the one-way purity $p_2$. In each panel, the dotted lines display the corresponding two-way-quantities being $c_2$ or $p_2$. It is shown that the purity obtained from the multi-run-procedure is more balanced than that from the single run. For FOF this leads also to a more balanced completeness.\label{fig:single_vs_multi}}
\end{figure}

\begin{figure}
		\epsscale{1}\plotone{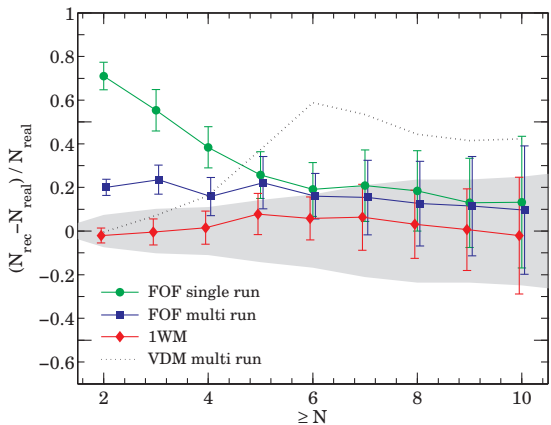}
		\figcaption{Relative abundance of reconstructed groups as compared with the real groups as a function of richness $N$. The green line shows the mean relative abundance of single-run FOF groups, the blue line the mean relative abundance of multi-run FOF groups, and the red line the mean relative abundance of one-way-matched groups. The errorbars always exhibit the scatter among the 24 mocks. The gray shaded region displays the spread of the relative abundance of real groups among the 24 mocks (i.e. cosmic variance plus shot-noise). For $N \lesssim 6$ the number of multi-run FOF groups is slightly too high and exceeds the margin of cosmic variance while the abundance of the one-way-matched groups is well within the region dominated by cosmic variance. For comparison the relative abundance of multi-run-VDM groups is shown as well (black dotted line).\label{fig:relative_numbers}}
\end{figure}

Since $c_1(N)$, $c_2(N)$, etc. are functions of richness $N$, one might wonder how the best $g_1(N)$-sets for different $N$ are distributed in the $c_1(8)$-$p_1(8)$-plane. This is shown in Figure \ref{fig:parameter_scatter} by the labeled points where the labels denote the corresponding $N$. For FOF, the best $g_1(8)$-set is optimal for all $N\geq 8$ as well, while for $N<8$ the optimal $g_1(N)$-sets reside at lower completeness. For VDM this is less obvious, but at least for $N\geq 10$ the best $g_1(N)$-sets seem to converge. In any case, it is clear that it is not possible to simultaneously optimize $g_1(N)$ for all $N$ with a single parameter-set. If the parameter-set is optimized for groups with $N\geq 8$, the resulting group catalogue is very complete for groups with $N<8$ but the purity starts to decrease severely for $N < 5$, and a lot of spurious small groups enter the catalogue (see Figures \ref{fig:single_vs_multi} and \ref{fig:relative_numbers}). Since around $\simeq 80 \%$ of the groups have $N < 5$, this is unsatisfactory.

This suggests that the groupfinder should be run several times with different parameter-sets, each time optimized for a different richness range.  This is analogous to the ``hot-cold'' double pass approach often used with image detection algorithms such as SExtractor.  We will refer to this approach as the ``multi-run procedure'', and it was implemented it as follows:
\begin{enumerate}
\item The parameter-set is optimized for the range $N\geq 6$, the groupfinder is run, and only those groups that are in this richness range are kept in the group catalogue.
\item The parameter-set is then optimized for groups with $N=5$, the groupfinder is run again, and only groups with $N=5$ that are not yet detected in the first step are added to the group catalogue.
\item Repeat the previous step for $N=4$.
\item Repeat the previous step for $N=3$.
\item Repeat the previous step for $N=2$. 
\end{enumerate}
In each step, only those groups are accepted which have not been found in an earlier step. It is better to work down in richness because the richer groups are more easily detected. The optimal parameter set in each step is basically just the best $g_1$-set for the corresponding richness range. However, particularly in the first step, also other choices are possible. In fact, for VDM, we have chosen a special set for the first step since the best $g_1(6)$-set proved to be by no means optimal for $N\gtrsim 8$. Table~\ref{tab:optimal-parameter-sets} gives the optimal parameter-sets for FOF and VDM. Since there are some degeneracies between the parameters, there are no simple trends from step 1 to step 5 for the single parameters.

Figure \ref{fig:single_vs_multi} shows how the multi-run-procedure compares to the single run best $g_1(8)$-set. In the case of FOF, the completeness has slightly decreased for $N \lesssim 5$ compared to the single run, but the high completeness of the single run in this richness range comes at the cost of a low purity. In fact, for the multi-run, the purity has increased for $N \lesssim 5$, and has become almost constant for all richness classes. Thus, the overall behaviour of the completeness and purity is now more balanced.

For VDM, we observe a similar trend. Here, it is particularly evident that in a single run, even if optimized for $N\geq 8$, the completeness decreases for $N \gtrsim 6$. The multi-run-procedure can correct for this and still increase the purity for small groups.

While the overall statistics of the two multi-run-catalogues are similar, there are some minor differences: The overall behaviour of the completeness and purity as a function of $N$ seems to be more balanced for FOF. Also the ratios $c_2/c_1$ and $p_2/p_1$ are more balanced for FOF, while for VDM, $c_2/c_1$ increases and $p_2/p_1$ decreases toward higher $N$. On the other hand, the total number of groups found with FOF is too high for $N \lesssim 3$, while for VDM, the number of reconstructed groups is too high for $N \gtrsim 5$ (Figure \ref{fig:relative_numbers}). All things considered, the multi-run-procedure works better with FOF than with VDM.

\subsubsection{Combining FOF and VDM} \label{sec:merging}

\begin{figure}
		\epsscale{1}\plotone{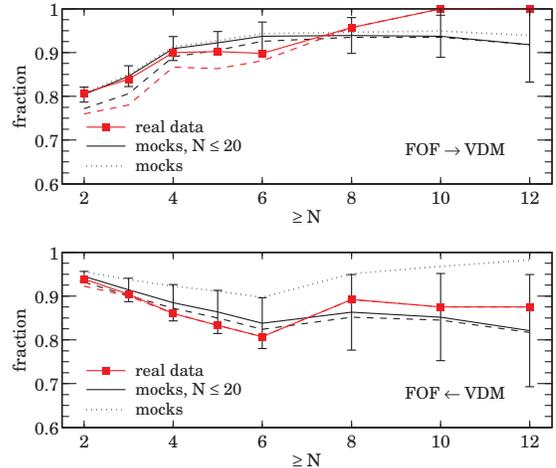}
		\figcaption{Comparison between the multi-run FOF and the multi-run VDM catalogues. The upper panel shows the fraction of FOF groups associated to VDM-groups. The red corresponds to the real data 10k group catalogue, whereas the solid line designates one-way-matches, and the dotted line two-way-matches. The black solid line corresponds to the mean fraction of associations in the mocks, if groups with $N>20$ are omitted, and the error bars exhibit the scatter among the 24 mocks. The black dotted line shows the same, if all groups are taken into account. In the lower panel, the symbols have identical meaning but exhibit the fraction of VDM groups associated to FOF groups.\label{fig:fof_vdm_association}}
\end{figure}

With the FOF and VDM multi-run group catalogues, there are now two catalogues available, obtained by different algorithms, and exhibiting similar purity and completeness. A comparison of the two catalogues on a group-to-group basis is shown in Figure \ref{fig:fof_vdm_association}.
The red lines show the result for the real 10k sample. An FOF group with $N\geq 2$ has a probability of being associated with a VDM group of $\sim 80\%$, increasing roughly linear with $N$ until it reaches $100\%$ for $N \geq 10$. On the other hand, the probability of any VDM group being associated with a FOF group is greater than $\sim 80\%$, and even higher than $\sim 90 \%$ for $N \geq 8$. The reason that for $N \lesssim 4$ the VDM groups have a higher probability of being associated with the FOF groups than vice versa is due to the excess production of small groups in the FOF catalogue. Furthermore, note that whenever a group with $N \geq 6$ has an associated group this association is a two-way-association. Thus, the two catalogues, though not identical, contain mainly the same structures. Moreover, the real data agree very well with the mocks (black solid lines), if groups with $N>20$ in the mocks are omitted (in the mocks there are too many of them, see \S~\ref{seq:n_of_N}). This shows that the groupfinders work indeed comparably on the real data as they do on the mocks.

Is there a way to combine the information in the two catalogues in order to obtain a single optimal group catalogue? It seems natural to consider those group galaxies that were recovered by both groupfinders. We introduce a ``galaxy purity parameter'' (GAP) for each galaxy.  The GAP is a flag indicating if a certain group galaxy is contained simultaneously in both catalogues. For a certain FOF group galaxy it is defined as follows:
\begin{itemize}
\item If there is no VDM group containing this galaxy, it gets a GAP equal to 0.
\item If it is also contained in a VDM group, and the FOF group has a one-way-match to this VDM group or this VDM group exhibits a one-way-match to the FOF group, the galaxy gets a GAP equal to 1.
\item If it is contained in a VDM group, and the FOF group has a two-way-match to this VDM group, the galaxy gets a GAP of 2.
\end{itemize}
Thus, we expect that the higher the GAP for a galaxy, the more reliable the detection, and the higher the probability that this galaxy is a real group galaxy and not an artefact introduced by one of the groupfinders. The GAP is a useful flag for excluding uncertain group members if needed, and defines more clearly the reliable core of a group.

Then, we can define two sub-sets, or ``sub-catalogues'', of the basic FOF catalogue. The ``one-way-matched'' (1WM) sub-catalogue contains only FOF group galaxies with a $\rm{GAP}\geq 1$. In a similar way, the ``two-way-matched'' sub-catalogue contains only group galaxies with a GAP = 2, i.e. all galaxies with a GAP $\leq 1$ become field galaxies. Note that we have defined the GAP, and thus the 1WM and 2WM, based on the FOF groups. They could, of course, also be defined based on the VDM groups. However, to obtain a single optimal group catalogue, we have to choose between FOF and VDM. As discussed in the last section, though the multi-run catalogues obtained by these groupfinders exhibit similar statistics, some (minor) properties are overall better for FOF. So we have decided the FOF catalogue to be the basic catalogue. The VDM catalogue, by contrast, is therefore only used to determine the GAPs of FOF group members. Since the two sub-catalogues preserve the group structure of the basic FOF catalogue, this set of three group catalogues can be presented as one single big catalogue with the GAP flags to indicate the increasing purity.

\subsection{Results on the mocks}

In this section, we will summarize our findings and give a detailed statistical description of the FOF catalogue with its two sub-catalogues (1WM and 2WM).

\begin{figure*}
		\epsscale{1}\plotone{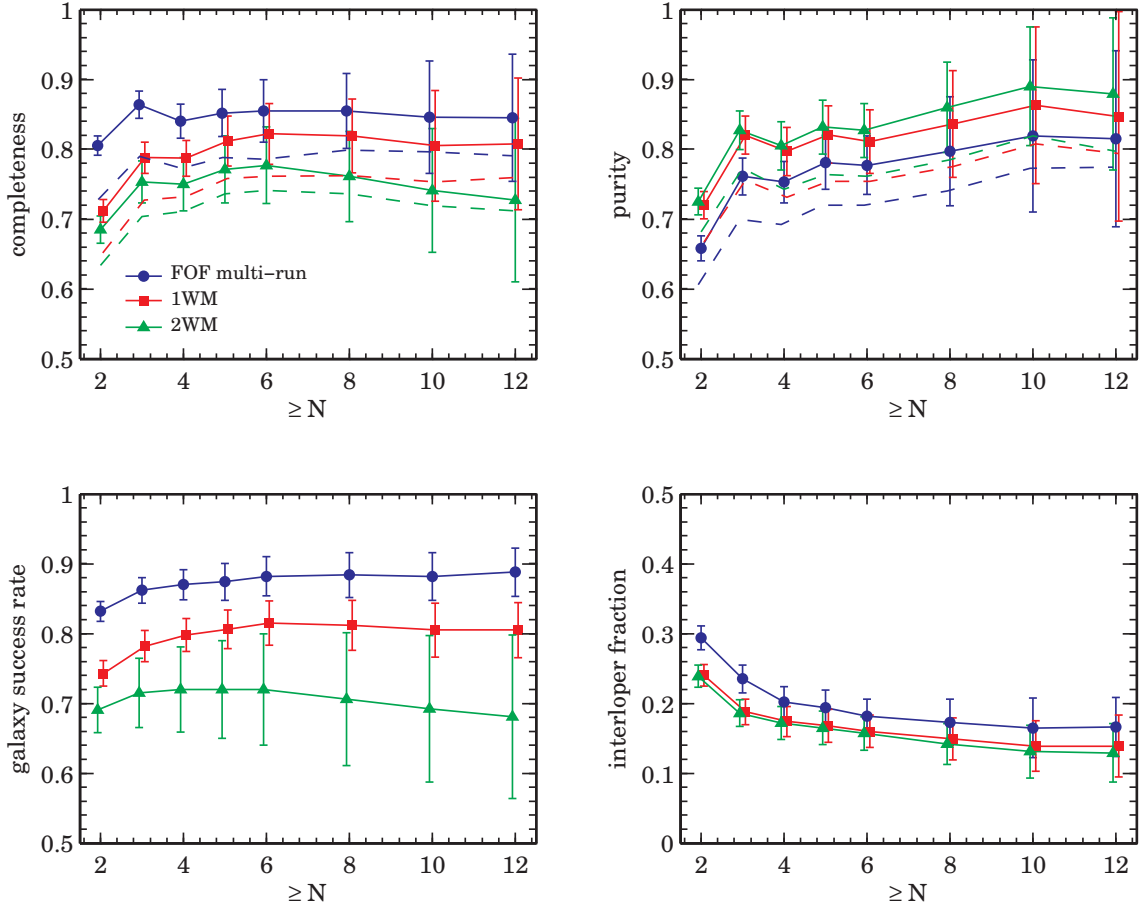}
		\figcaption{Statistics of the FOF and its two sub-catalogues, WM1 and WM2, as function of richness $N$. For all panels, blue refers to the FOF groups, red to the one-way-matched groups and green to the two-way-matched groups. The errorbars show the scatter among the 24 mocks. The upper left panel exhibits completeness and the upper right panel purity. The solid lines correspond to $c_1$ and $p_1$ respectively and the dashed line $c_2$ and $p_2$ respectively. The lower right panel shows the galaxy success rate $S_{\rm gal}$ and the lower left panel the interloper fraction $f_{\rm I}$.\label{fig:statistics}}
\end{figure*}

The statistics of the merged catalogues in comparison with the reference FOF catalogues is shown in Figure \ref{fig:statistics} and for $N\geq 5$ in Table \ref{tab:statistics}.
\begin{deluxetable}{ccccccc}
\tablewidth{0.4\textwidth}
\tablecaption{Catalogue statistics for $N\geq 5$}
\tablehead{
  \colhead{catalogue} &
  \colhead{$c_1$\tablenotemark{a}} &
  \colhead{$p_1$\tablenotemark{b}} &
  \colhead{$c_2/c_1$} &
  \colhead{$p_2/p_1$} &
  \colhead{$S_{\rm gal}$\tablenotemark{c}} &
  \colhead{$f_{\rm I}$\tablenotemark{d}} 
  }
\startdata
FOF & 0.85 & 0.78 & 0.92 & 0.92 & 0.87 & 0.19  \\
1WM & 0.81 & 0.82 & 0.93 & 0.92 & 0.81 & 0.17  \\
2WM & 0.77 & 0.83 & 0.95 & 0.92 & 0.72 & 0.17  
\enddata

\tablenotetext{a}{One-way completeness}
\tablenotetext{b}{One-way purity}
\tablenotetext{c}{Galaxy success rate}
\tablenotetext{d}{Interloper fraction}

\tablecomments{The precise definitions of the statistical quantities are given in \S~\ref{sec:statistics}.}

\label{tab:statistics}
\end{deluxetable}
The lines exhibit the mean among the 24 mocks and the errorbars their scatter. The FOF basis catalogue has a completeness $c_1 \simeq 0.85$ almost not depending on the richness $N$ and a purity $p_1 \simeq 0.78$ only weakly depending on $N$. Only for $N=2$ there is a significant decrease in both completeness and purity. The corresponding statistics for the 1WM and 2WM sub-catalogues have almost identical dependences on $N$ but, as expected, their $c_1$ is lower and $p_1$ higher.

It can be seen that the gain of the 2WM catalogue compared to 1WM in terms of both purity or interloper fraction is much smaller than the deficit in terms of completeness and galaxy success rate. This indicates that by keeping only group galaxies with a GAP = 2, many real group galaxies are removed, but only a relatively small number of interlopers are eliminated. By contrast, the gain in purity of the 1WM with respect to the reference FOF is quite comparable to the associated decrease in completeness. Thus, while the 1WM catalogue is a useful construction, little is gained by the more restrictive 2WM catalogue.  In the remainder of this paper, we will mainly refer to the FOF and its 1WM sub-catalogue.  We note that not only do the ratios $c_2/c_1$ and $p_2/p_1$ behave well as a function of $N$ for the three catalogues, but also $c_2/c_1 \simeq p_2/p_1$. This means that the contributions of over-merging and fragmentation are not only small, but are also well-balanced.

\begin{figure}
		\epsscale{1}\plotone{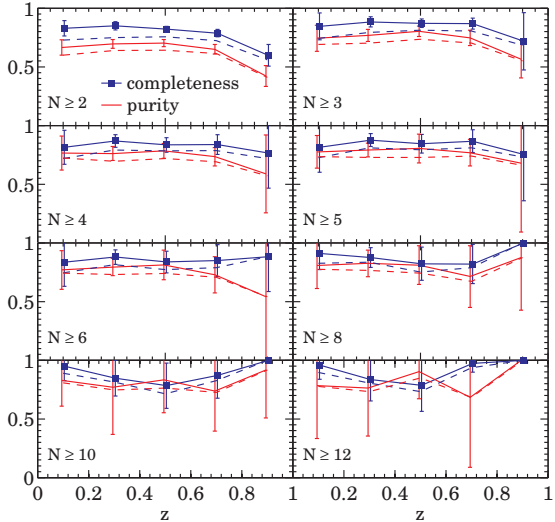}
		\figcaption{Completeness and purity of the FOF groups as a function of redshift for 8 different richness classes. In each panel, the blue solid line corresponds to the mean $c_1$-completeness and the red solid line to the mean $p_1$-purity, whereas the errorbars exhibit the scatter among the 24 mocks. The dashed lines are for the corresponding 2-way-quantities, respectively. The richness class $N$ is indicated in each panel.\label{fig:statistics_z}}
\end{figure}

So far, we have considered the statistics averaged over the whole redshift range, i.e. $0.1 \lesssim z \lesssim 1$. In Figure \ref{fig:statistics_z}, the completeness (blue line) and the purity (red line) of the FOF catalogue are shown as functions of redshift for several richness classes $N$.
The curves are consistent with a relatively constant completeness and purity with redshift. Only the highest redshift bins for $N \lesssim 4$ show possibly a slight decrease. This emphasizes further the robustness of our catalogue.
\begin{figure}
		\epsscale{1}\plotone{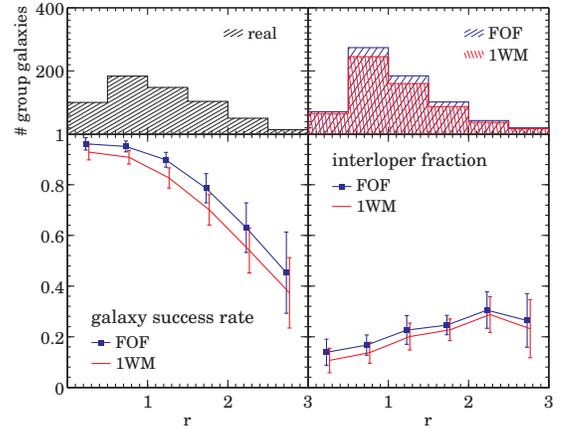}
		\figcaption{Behaviour of the galaxy success rate $S_{\rm gal}$ and the interloper fraction $f_{\rm I}$ as a function of the normalized projected  distance $r$ from the group centers, where $r$ is defined in Equation~\ref{eq:r}. The left lower panel shows the galaxy success rate $S_{\rm gal}$, where the blue line corresponds to the FOF and the red line to the 1WM catalogue. The left upper panel shows the distribution of real group galaxies as a function of separation from the cluster centers. It is clear that at $r \lesssim 1.5$, where most real group galaxies reside, $S_{\rm gal}$ is $\gtrsim 0.9$ for FOF groups and only slightly lower for 1WM groups. The right lower panel exhibits interloper fraction $f_{\rm I}$ and the right upper panel the distribution of galaxies in reconstructed groups as a function of $r$, whereas blue corresponds to FOF and red to 1WM.\label{fig:completeness_radius}}
\end{figure}
Figure \ref{fig:completeness_radius} shows how the galaxy success rate $S_{\rm gal}$ and the interloper fraction $f_{\rm I}$ behave as a function of the normalized projected distance from the group centers.

The distance variable $r$ is defined for each group galaxy as
\begin{eqnarray}
r = \sqrt{\left(\frac{\theta_{\rm ra}}{\Delta \theta_{\rm ra}}\right)^2 + \left(\frac{\theta_{\rm dec}}{\Delta \theta_{\rm dec}}\right)^2},\label{eq:r}
\end{eqnarray}
where $\theta_{\rm ra}$ is its separation from the group center in $\alpha$, and $\Delta \theta_{\rm ra}$ the second moment in $\alpha$ among all members of this group. Similar definitions hold for $\theta_{\rm dec}$ and $\Delta \theta_{\rm dec}$. Only groups with 3 or more members are taken into account, since for groups with only 2 members $r$ becomes meaningless.

The left lower panel shows the galaxy success rate $S_{\rm gal}$ as a function of $r$ from the real group centers. As one would expect, it increases toward the group centers. Fortunately, the group centers are also the region where most of the real group galaxies reside (left upper panel).  Note that $S_{\rm gal}$ can decrease, in principal, in two ways: First of all, by failing to identify certain real group galaxies in successfully detected real groups, and second, by failing to detect a real group at all. The small deficit in $S_{\rm gal}$ for $r \lesssim 1$ is due to the second reason, while the first reason becomes more important with increasing $r$.

In the right lower panel, the interloper fraction $f_{\rm I}$ is plotted as a function of $r$, where $r$ is now related to the centers of the reconstructed groups. As expected, the interloper fraction shows the opposite behaviour as a function of $r$. However, the difference in $f_{\rm I}$ between near and far galaxies from the group centers is less strong than for $S_{\rm gal}$. For small $r$, the most important contribution to $f_{\rm I}$ comes from spuriously detected groups with 3 members.

Finally, Figure \ref{fig:relative_numbers} shows the numbers of reconstructed groups relative to the number of real groups. As was already mentioned, the mean difference between the number of reconstructed FOF groups and the number of real groups exceeds the uncertainty expected by cosmic variance from mock to mock for $N \lesssim 5$, while the groups of the 1WM sub-catalogue are well within this region.

According to the statistics discussed in this paragraph, particularly in Figure\ref{tab:statistics}, it became clear that the FOF group catalogue along with its 1WM sub-cataloge has the potential to be useful for many different applications such as galaxy evolution studies, group statistics, or gravitational lensing. For example, if one aims to study the evolution of galaxies in groups, a high purity and a low interloper fraction are desirable, so the 1WM catalogue is probably appropriate. On the other hand, in order to have a relative pure sample of field galaxies, galaxies not contained in the basic FOF catalogue should be selected. Generally, it holds that whenever small groups, number of groups, or purity of the group sample is important, the 1WM catalogue is to be preferred to the FOF catalogue.

\subsection{Comparison with DEEP2}

For DEEP2, Gerke et al. (2005) optimized their VDM groupfinder in order to obtain the correct number of reconstructed groups $N_{\rm gr}^{\rm rec}(\sigma,z)$ as a function of velocity dispersion $\sigma$ and redshift. As result, they present two group catalogues: an ``optimal'' catalogue and one with maximized purity. Since Gerke et al. (2005) did not treat completeness and purity as a function of richness $N$, all their statistics correspond to $N\geq 2$.

The statistics for their optimal parameter set are $c_1 = 0.782 \pm 0.006$, $p_1 = 0.545 \pm 0.005$, $S_{\rm gal} = 0.786$, and $f_{\rm I} = 0.458 \pm 0.004$. The ratios between the two-way and the one-way-quantities are therefore $c_2/c_1 = 0.919$ and $p_2/p_1 = 0.987$. So in comparison with our own FOF $N\geq 2$ statistics, their completeness $c_1$ and galaxy success rate $S_{\rm gal}$ are $\sim 3 \%$ and $\sim 6 \%$ lower, respectively, while their purity $p_1$ is $\sim 17 \%$ lower, and their interloper fraction $f_{\rm I}$ $\sim 56 \%$ higher.

We conclude that, compared with the DEEP2 ``optimal'' group catalogue, the performance of our FOF group catalogue is very high. Moreover, it would be very interesting to compare the statistics for the higher richness classes as well. Since Gerke et al. (2005) optimized their catalogue using all groups with $N\geq 2$ their catalogue should be optimal regarding the $N \geq 2$ statistics. But, in contrast to a multi-run catalogue, this might not be the case for the higher richness statistics, since the $N\geq 2$ statistics are actually dominated by 2-member groups being by far the most abundant. This suggests that the relative superiority of our FOF catalogue over the DEEP2 catalogue could be even higher for the higher richness classes.

\section{The real data 10k group catalogue}

In this section, the real data 10k group catalogue is presented. It is given by means of the Tables \ref{tab:groups} and \ref{tab:galaxies}.
Table \ref{tab:groups} is a list of all groups along with their properties, and Table \ref{tab:galaxies} is a list of all group galaxies. The group galaxies are associated to their group by means of the unique group-ID. The galaxy-IDs refer to the 10k catalogue published by S. J. Lilly et al. (2009, in preparation).
\begin{deluxetable*}{crcccrcl}
\tablewidth{0.6\textwidth} 
\tablecaption{Group catalogue (excerpt)}
\tablehead{
  \colhead{Group-ID} &
  \colhead{$N$\tablenotemark{a}} &
  \colhead{$\left\langle \alpha \right\rangle$} &
  \colhead{$\left\langle \delta \right\rangle$} &
  \colhead{$\left\langle z \right\rangle$} &
  \colhead{$\hat \sigma$\tablenotemark{b}} &
  \colhead{$M_{\rm fudge}$\tablenotemark{c}} &
  \colhead{GRP$_1$\tablenotemark{d}} \\
  \colhead{} &
  \colhead{} &
  \colhead{(deg)} &
  \colhead{(deg)} &
  \colhead{} &
  \colhead{(km/s)} &
  \colhead{($M_\odot$)} &
  \colhead{} 
}
\startdata
0  & 10 & 150.0087 & 2.0287 & 0.0788 & 409 & 8.20e12 & 0.9 \\
1  &  7 & 149.4817 & 2.5073 & 0.0919 & 393 & 9.58e12 & 1 \\
2  & 20 & 150.3004 & 2.4489 & 0.1231 & 442 & 6.09e13 & 0.95 \\
3  &  6 & 150.3444 & 2.1544 & 0.1222 & 0   & 1.13e13 & 1 \\
4  & 14 & 149.8568 & 1.8151 & 0.1243 & 532 & 1.92e13 & 0.93 \\
5  &  6 & 149.7238 & 2.399  & 0.1252 & 0   & 1.23e13 & 1 \\
6  &  6 & 150.2824 & 2.1531 & 0.1686 & 69  & 5.98e12 & 0.5 \\
7  &  9 & 150.078  & 2.2136 & 0.1865 & 242 & 2.18e13 & 1 \\
8  & 12 & 150.1122 & 2.3564 & 0.2208 & 596 & 2.71e13 & 0.92 \\
9  & 10 & 150.4526 & 2.6799 & 0.2179 & 768 & 2.83e13 & 1 \\
10 &  6 & 150.2371 & 1.9404 & 0.2188 & 238 & 1.13e13 & 1 \\
11 &  6 & 150.2304 & 2.5608 & 0.2207 & 530 & 2.29e13 & 0 \\
12 &  6 & 150.1636 & 2.0342 & 0.2208 & 355 & 1.35e13 & 1 \\
13 &  8 & 150.2494 & 2.6574 & 0.2672 & 403 & 3.85e13 & 1
\enddata
\tablecomments{The full table is available in electronic form.}
\tablenotetext{a}{Observed richness}
\tablenotetext{b}{Velocity dispersion (see \S~\ref{sec:velocity_dispersion})}
\tablenotetext{c}{Virial mass of the DM halo (see \S~\ref{sec:mass})}
\tablenotetext{d}{Group purity parameter (GPR$_1$) (see \S~\ref{sec:group_purity_parameter})}

\label{tab:groups}
\end{deluxetable*}
\begin{deluxetable}{ccrcccc}
\tablecaption{Group galaxies (excerpt)}
\tablehead{
  \colhead{Galaxy-ID} &
  \colhead{Group-ID} &
  \colhead{$N$} &
  \colhead{$\alpha$} &
  \colhead{$\delta$} &
  \colhead{$z$} &
  \colhead{GAP\tablenotemark{a}}  \\
  \colhead{} &
  \colhead{} &
  \colhead{} &
  \colhead{(deg)} &
  \colhead{(deg)} &
  \colhead{} &
  \colhead{}
}
\startdata
818787 & 0 & 10 & 150.0605 & 2.0067 & 0.0785 & 2 \\
818888 & 0 & 10 & 150.0365 & 2.0249 & 0.0794 & 2 \\
818934 & 0 & 10 & 150.0241 & 1.9687 & 0.0779 & 2 \\
818935 & 0 & 10 & 150.0239 & 2.0727 & 0.0779 & 2 \\
818982 & 0 & 10 & 150.0134 & 2.0296 & 0.0791 & 2 \\
819035 & 0 & 10 & 149.9989 & 1.9858 & 0.0805 & 0 \\
819041 & 0 & 10 & 149.9984 & 2.0351 & 0.0789 & 2 \\
819060 & 0 & 10 & 149.9912 & 1.9912 & 0.0797 & 2 \\
819118 & 0 & 10 & 149.9724 & 2.1054 & 0.0781 & 2 \\
819133 & 0 & 10 & 149.9681 & 2.0673 & 0.0779 & 2 \\
842033 & 1 & 7 & 149.4897 & 2.5164 & 0.0913 & 2 \\
842048 & 1 & 7 & 149.4844 & 2.4991 & 0.0907 & 2 \\
842049 & 1 & 7 & 149.4839 & 2.5211 & 0.0915 & 2
\enddata
\tablecomments{The full table is available in electronic form.}
\tablenotetext{a}{Galaxy purity parameter (see \S~\ref{sec:merging})}

\label{tab:galaxies}
\end{deluxetable}

\subsection{Group purity parameter} \label{sec:group_purity_parameter}

Since we are presenting the FOF catalogue along with its two sub-catalogues, defined by the GAP parameter in the final column, any group property can, in principle, be calculated for all three catalogues. For instance, it is possible to assign to each group three observed richnesses $N$. To avoid confusion and to keep the discussion simple, all group properties given in Table \ref{tab:groups} correspond to the basic FOF catalogue. In order to quantify the number of 1WM galaxies in a certain group, we introduce the group purity parameter (GRP$_i$) for $i=1,2$, defined by the fraction of FOF members having a galaxy purity parameter GAP $\geq i$. For $i=1$ this is the fraction of FOF members that are also 1WM members, and for $i=2$ that are also 2WM members. Note that if the GRP$_1$ is zero, then there is no association between the FOF group and a VDM group.

The statistics of the number of groups and the GRP$_1$ are summarized in Table~\ref{tab:GRP}. The basic FOF catalogue contains 800 groups with $N \geq 2$, 102 groups with $N \geq 5$, and 23 groups with $N \geq 8$. Over $80 \%$ of the groups with $N \geq 2$ have a GRP$_1$ greater than zero, i.e. these groups have at least one group galaxy that was independently recovered by both FOF and VDM. For the groups with $N \geq 5$, the number of groups with GRP$_1 > 0$ rises to $95 \%$, and for those with $N \geq 8$ it is $96 \%$ (22 out of 23).  Figure \ref{fig:fof_vdm_association} shows the comparison between the real data FOF and VDM catalogues.
\begin{deluxetable}{crcc}
\tablewidth{0.35\textwidth}
\tablecaption{Number of groups}
\tablehead{
  \colhead{$\geq N$} &
  \colhead{$N_{\rm gr}$\tablenotemark{a}} &
  \colhead{$f({\rm GRP}_1>0)$\tablenotemark{b}} &
  \colhead{$<\rm{GRP}_1>$}
}

\startdata
2 &  800 & 0.82 &  0.80  \\
3 &  286 & 0.86 &  0.81  \\
4 &  150 & 0.95 &  0.88  \\
5 &  102 & 0.95 &  0.88  \\
6 &   59 & 0.95 &  0.87  \\
7 &   36 & 0.94 &  0.86  \\
8 &   23 & 0.96 &  0.88  \\
9 &   17 & 1.00 &  0.93  
\enddata

\tablenotetext{a}{Number of groups}
\tablenotetext{b}{Fraction of groups with a group purity parameter (GPR$_1$) larger than zero.}

\label{tab:GRP}
\end{deluxetable}

\begin{figure}
		\epsscale{1}\plotone{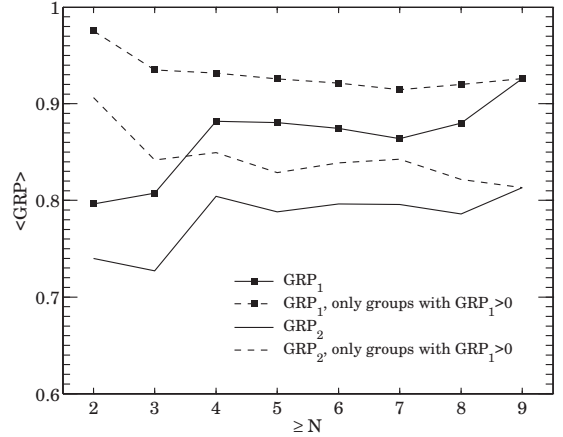}
		\figcaption{Mean GRP as a function of observed richness $N$. The blue solid line shows the GRP$_1$ and the red solid line the GRP$_2$. The dashed lines show the corresponding GRP$_i$, $i=1,2$, by taking into account only groups with a non zero GRP$_1$.\label{fig:mean_GRP}}
\end{figure}

The mean GRP as a function of richness $N$ is given in Figure \ref{fig:mean_GRP}.
The blue solid line shows the mean GRP$_1$ taking into account all groups with $\geq N$. There is a slight and noisy rise from about 0.8 for $N\geq 2$ to 0.9 for $N \geq 9$ due to the fact that the fraction of groups with GRP$_1$ equals zero is slightly bigger for smaller $N$. On the other hand, the dashed blue line shows the mean GRP$_1$ taking into account only groups with a non zero GRP$_1$, i.e. only groups simultaneously found by both groupfinder. For these groups, the GRP$_1$ is slightly decreasing since for bigger groups it becomes easier for the two groupfinders to disagree on one or two galaxies in the outskirts of the group. The red lines in Figure \ref{fig:mean_GRP} show the same quantities for GRP$_2$.

\subsection{Corrected richness $N_{\rm corr}$}

\begin{figure}
		\epsscale{1}\plotone{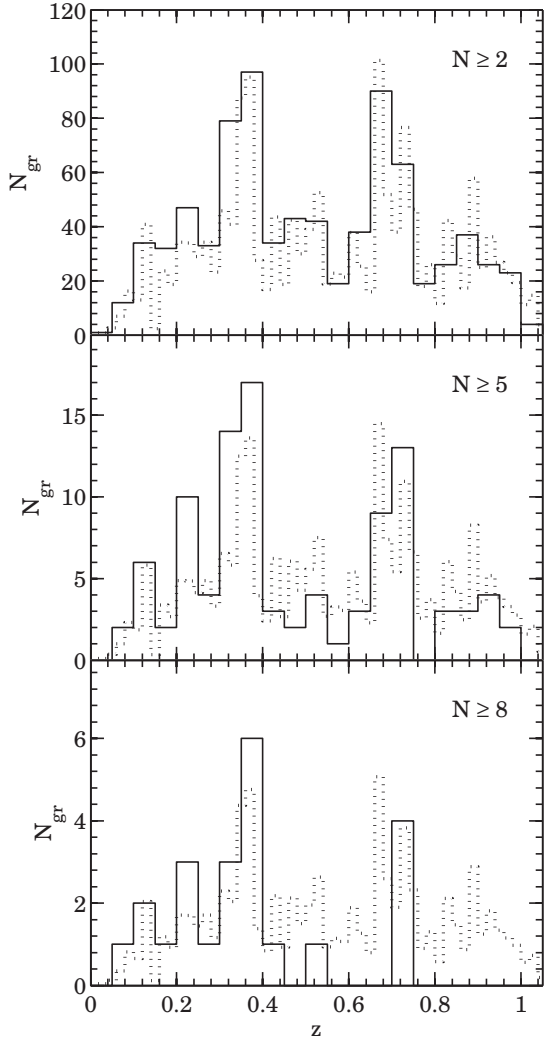}
		\figcaption{Number of groups as a function of redshift for different richness classes $N$. The top panel shows the number of groups $N_{\rm gr}$ for groups with $N \geq 2$, the middle panel for $N \geq 5$, and the bottom panel for $N \geq 8$. The red dotted line shows the number of galaxies $N_{\rm gal}$ for the galaxy sample scaled down for comparison with the groups. It is obvious that the distribution of groups follows the distribution of galaxies.\label{fig:n_of_z}}
\end{figure}

The distribution of FOF groups as a function of redshift for three richness classes $N$ is shown in Figure \ref{fig:n_of_z}.
Comparing the black histograms (groups) with the red dashed lines (all galaxies) it is clear that the number of groups at a given redshift scales with the number of galaxies at the same redshift. This is basically true for all richness classes although for the richest $N \geq 8$ there is a lack of groups at redshifts $z \gtrsim 0.5$. In the framework of the hierarchical cold dark matter (CDM) structure formation scenario we expect the cluster mass function to grow with time (for a review see Voit 2005). This growth should be reflected in the decrease of the number of groups of a given richness with redshift. 

In order to address this question, it is necessary to correct the observed richness of a cluster to produce an intrinsic richness that is redshift-independent.  We therefore introduce the corrected richness $N_{\rm corr}$, correcting the observed richness $N$ for spatial sampling rate and redshift success rate, and considering for each group only the number of members brighter than a given absolute magnitude limit $M_{\rm b,lim}(z)$, i.e for each group
\begin{eqnarray}
N_{\rm corr}(M_{\rm b,lim}(z)) = \sum_{i} \frac{1}{C_{\alpha\delta,i}}\frac{1}{C_{{\rm rsr},i}},
\end{eqnarray}
where the sum is over the members of the group with $M_b \leq M_{b,\rm lim}(z)$, and $C_{\alpha\delta,i}$ and $C_{\rm{rsr},i}$ are the sampling rate and the redshift success rate, respectively, for the galaxy $i$. The redshift dependence of the absolute magnitude limit is always taken to be $M_{b,\rm lim}(z) = M_{b,\rm lim}-z$, whereas the subtraction of the redshift is to account approximately for the luminosity evolution of the galaxies. So $N_{\rm corr}$ can simply be characterized by $M_{b,\rm lim}$ being the absolute magnitude limit at redshift zero. Absolute magnitudes were obtained by means of standard multicolor spectral energy distribution (SED) fitting using an uptdated version of the ZEBRA code (Feldmann et al. 2006, P. Oesch et al. 2009 in preparation).

\begin{figure}
		\epsscale{1}\plotone{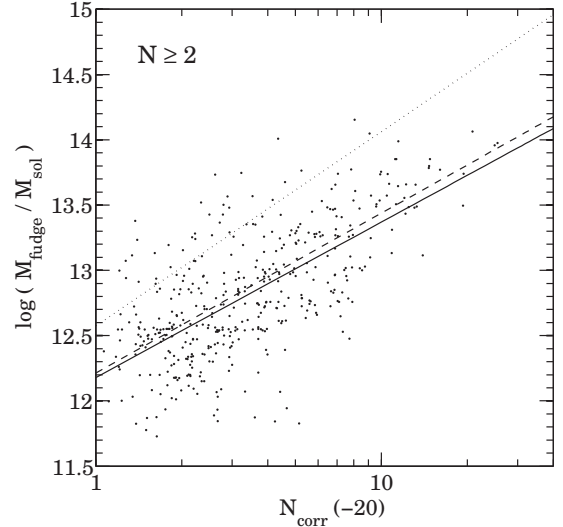}
		\figcaption{Correlation between $N_{\rm corr}(-20)$ and $M_{\rm fudge}$ (see \S~\ref{sec:mass}) for the 10k groups. Shown are all groups region having a redshift $z < 0.8$, so that the sample is volume limited for $M_{\rm b,lim}(z) = -20 -z$. The solid line is a linear regression through the points, and the dashed line is the same quantity for the reconstructed groups in the mocks not shown here. The dotted line exhibits the linear regression for the $N_{\rm true}(-20)$-$M$-relation for the real groups in the mocks. Taking into account the overestimation of $N_{\rm corr}$ of about $50\%-100 \%$ (see the text), the dotted curve can be reconciled with the solid one.\label{fig:n_corr_mass}}
\end{figure}

If we denote the actual number of group members brighter than $M_{\rm b,lim}(z)$ in a real group in the mocks (without the spatial or redshift sampling rates) by $N_{\rm true}(M_{\rm b,lim})$, then we find that, for reconstructed groups exhibiting a two-way-match to real groups, the estimated $N_{\rm corr}(-20)$ exhibit a relatively large scatter ($\pm 50 \%$) compared with $N_{\rm true}(-20)$ of the corresponding real groups. Furthermore, $N_{\rm corr}(-20)$ on average overestimates $N_{\rm true}(-20)$ by about $50-100 \%$ depending on $N$. This is because (1) for most groups we are in the low number regime, (2) the sampling rate in the 10k sample is rather low (so the corrections are big and noisy), (3) groups with no galaxy brighter than $M_{\rm b,lim}(z)$ cannot be corrected for sampling rate at all, and (4) the reconstructed groups are affected by interlopers. One should therefore be cautious in interpreting $N_{\rm corr}$ as the actual richness of the groups. Nevertheless, $N_{\rm corr}(-20)$ shows a relatively tight correlation with the estimated halo mass $M_{\rm fudge}$ (see \S~\ref{sec:mass}), even for $N \geq 2$, as is shown in Figure~\ref{fig:n_corr_mass} and thus is still a useful quantity.
The analysis of the redshift distribution for groups with a given $N_{\rm corr}(-20)$ is performed in \S~\ref{sec:fraction_of_galaxies_in_groups}.

\subsection{Velocity dispersion estimation} \label{sec:velocity_dispersion}

The corrected richness $N_{\rm corr}$ discussed in the last section is probably the simplest and most straightforward characterization of a group. However, there are other characterizations of groups which may be more directly useful from a physical point of view such as velocity dispersion $\sigma$ or dynamical mass $M$. Since most of our groups have richness $N \lesssim 10$, we are in a low number regime, where the estimation of both velocity dispersion and mass is non-trivial.

According to Beers, Flynn \& Gebhardt (1990) the best estimators for velocity dispersion in groups with few members are the gapper estimator and the simple standard deviation. On the other hand, the biweight estimator seems to work very well on a large range of richness classes $N$ except for $N \lesssim 20$ where its performance is lower but still sufficient. For comparison, we have implemented all three estimators and none of them is significantly superior to the others when applied to the mocks, and so we will stick to the gapper estimator since it is the most commonly used among the three.

The implementation for a group with $N$ members is as follows: First of all, for each group member $i$ we computed the redshift difference $dz_i$ in respect to the mean group redshift $z_{\rm gr}$. Then these redshift differences were converted into velocities by
\begin{eqnarray}
dv_i = c\;dz_i/(1+z_{\rm gr})
\end{eqnarray}
with $c$ the speed of light. Then after sorting the velocities $dv_i$ in ascending order, the gapper estimate is given by
\begin{eqnarray}
\sigma_{\rm gap} = \frac{\sqrt{\pi}}{N(N-1)}\sum_{i=1}^{N-1} w_i g_i,
\end{eqnarray}
whereas the weights $w_i$ and the gaps $g_i$ are defined by
\begin{eqnarray}
w_i &=& i(N-1)\\
g_i &=& dv_{i+1}-dv_i
\end{eqnarray}
for $i=1,\ldots,N-1$. However in order to have a realistic estimate of the velocity dispersion of our group we have to correct $\sigma_{\rm gap}$ for our redshift uncertainty $\sigma_{\rm v}$ of roughly $100 \rm{\ km\;s^{-1}}$. This is done by
\begin{eqnarray}
\hat\sigma = \sqrt{3}\sqrt{\sigma_{\rm gap}^2 - \sigma_{\rm v}^2},\label{eq:subtraction}
\end{eqnarray}
where $\hat\sigma$ is the final estimate of the velocity dispersion $\sigma$. The factor $\sqrt{3}$ converts the line of sight velocity dispersion to the 3D velocity dispersion. If $\sigma_{\rm v}$ is larger than $\sigma_{\rm gap}$ we set $\hat\sigma$ formally to zero.

Since the COSMOS lightcones (Kitzbichler \& White 2007) provide only the ``virial velocity''\footnote{In the COSMOS lightcones, the virial velocity is simply defined by $v_{\rm vir} = \sqrt{G M_{200}/r_{200}}$, whereas $G$ is the gravitational constant, and $M_{200}$ and $r_{200}$ are the virial mass and the virial radius, respectively, related by $M_{200} = 4/3 \pi r_{200}^3 200 \rho_{\rm c}(z)$ with $\rho_{\rm c}(z)$ the critical density of the universe at the redshift of the halo.} $v_{\rm vir}$ of the DM halos and not directly the ``velocity dispersion'' $\sigma$, we cannot precisely estimate the uncertainty of the estimated $\hat\sigma$ for a group. But comparing $\hat\sigma$ to $v_{\rm vir}$ should provide an upper limit to the uncertainty. To take into account the influence of interlopers on $\hat\sigma$, we considered the estimated velocity dispersion of reconstructed groups exhibiting a two-way-match with real groups. (Wrongly detected groups do not exhibit a meaningful velocity dispersion.)

\begin{figure}
		\epsscale{1}\plotone{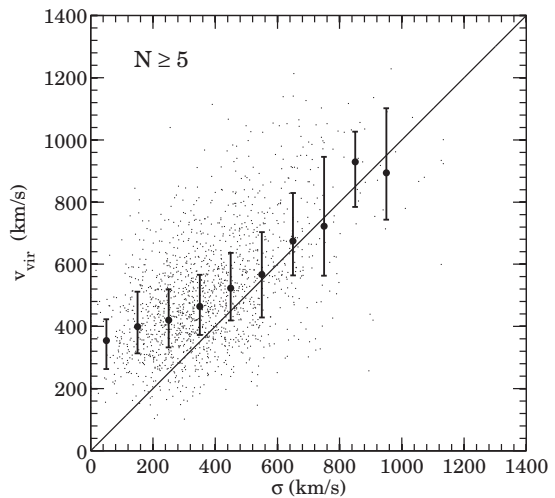}
		\figcaption{Correlation between the estimated velocity dispersions $\hat \sigma$ of groups with $N \geq 5$ and the virial velocities $v_{\rm vir}$ of the DM halos . Each point displays a reconstructed group exhibiting a two-way-match to a real group whose DM halo yields $v_{\rm vir}$. It is obvious that for $\hat \sigma \lesssim 350$ the estimated velocity dispersion is underestimating the virial velocity.\label{fig:velocity_dispersion}}
\end{figure}

We find that for $N \geq 5$ the ratio between the median virial velocity $v_{\rm vir}$ and $\hat\sigma$ remains roughly constant for $\hat\sigma \gtrsim 350\ \rm{km\ s^{-1}}$, and exhibits an error of about $25 \%$ (upper and lower quartile) (Figure~\ref{fig:velocity_dispersion}).
Note that the estimated $\hat\sigma$ do not need to fall exactly on the $45^\circ$-line, since $\sigma$ and $v_{\rm vir}$ are not exactly the same quantities. For $\hat\sigma \lesssim 350\ \rm{km\ s^{-1}}$ the estimated velocity dispersion $\hat\sigma$ is biased to lower values due to the subtraction in equation (\ref{eq:subtraction}). On the other hand, for $N < 5$, the correlation between $\hat\sigma$ and $v_{\rm vir}$ is very weak, so that the $\hat\sigma$ for these richness classes contains almost no information. Hence, we have decided to assign no estimated velocity dispersion to groups with $N<5$ in Table~\ref{tab:groups}.

Note that applying the velocity dispersion estimation to the real groups instead of the reconstructed groups does not significantly alter these results. Even estimating the velocity dispersion for the real groups in the 10k mocks taking into account all galaxies down to $r \leq 26$ still yields a scatter of about $10$ to $15 \%$.

\subsection{Estimation of dynamical mass}\label{sec:mass}

Estimating the dynamical mass of the underlying dark matter halo of a group is even more difficult than estimating the velocity dispersion. The simplest method for the estimation of dynamical mass is by using some form of the virial theorem. The standard relation is (e.g. Eke et al. 2004)
\begin{eqnarray}
\label{eq:virialtheorem}
\hat M = A \frac{\hat\sigma^2 r_\perp}{G},\label{eq:virial_mass}
\end{eqnarray}
where $A$ is a constant depending on the mass distribution of the halo (e.g. geometry, concentration, etc.), $\hat \sigma$ the estimated velocity dispersion, and $r_\perp$ is some estimate of its projected radius. Heisler, Tremaine \& Bahcall (1985) discuss four simple mass estimators, each being only a function of the projected distances and radial velocities of the group galaxies in respect to the group center. In applying them to the reconstructed groups in the mocks, none of them works substantially better than the simple relation in equation (\ref{eq:virialtheorem}) and all show a similar behaviour, so we consider only the standard virial theorem.

To use the estimator in Equation (\ref{eq:virialtheorem}), the constant of proportionality $A$ needs to be calibrated properly. Doing this with the mocks and using an appropriate estimation for the projected radius, we find a similar behaviour for the estimated masses like for the velocity dispersion. For $N \lesssim 5$ there is only a very weak correlation between the estimated mass and the actual mass of the underlying DM halo. For $N \gtrsim 5$ there is a correlation, but only for $\hat{M} \gtrsim 5 \times 10^{13}\ M_{\odot}$, and with an error (upper and lower quartiles) of roughly a factor of 2 with respect to the median. Since these mass estimates would be of relatively limited use and since the mocks are needed for calibration anyway, we have instead pursued another approach:

It turns out that using observed richness $N$, corrected for sampling and redshift success rate, and redshift $z$ as proxy for the mass and calibrating them with the mocks works rather well. The mass for a group with observed richness $N$ and redshift $z$ is then simply given by
\begin{eqnarray}\label{eq:fudge_mass}
M_{\rm fudge} =  \left\langle M_{\rm halo}(\tilde{N},z) \right\rangle,
\end{eqnarray}
where $M_{\rm halo}(\tilde{N},z)$ denotes the mass of a halo at redshift $z$ containing $\tilde{N}$ galaxies, $\tilde{N}$ is the observed richness of the group corrected for sampling and redshift success rate, and the angle brackets denote the average over the halos in the 24 mocks. We will denote this mass as ``fudge mass'' to indicate that it is calibrated with the mocks.

\begin{figure}
		\epsscale{1}\plotone{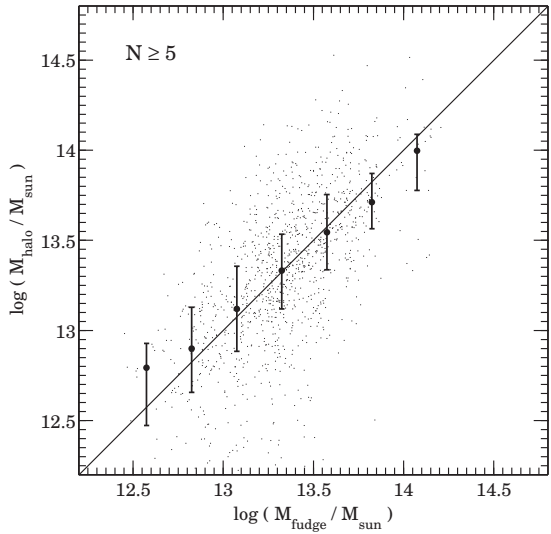}
		\figcaption{Correlation between the ``fudge mass'' $M_{\rm fudge}$ of groups with $N \geq 5$ and the virial mass $M_{\rm halo}$ of the corresponding DM halos. Each point displays a reconstructed group exhibiting a two-way-match to a real group to whose DM halo mass it is compared.\label{fig:mass}}
\end{figure}

The fudge masses of the reconstructed groups with $N\geq 5$ exhibiting a two-way-match to their real groups are shown in Figure~\ref{fig:mass}.
In contrast to the velocity dispersion estimates, there is no bias for small masses. The error for the masses (upper and lower quartiles) is about $50 \%$. Furthermore, the masses are also defined for small groups, whereas the upper quartile increases toward $100 \%$ for $N = 2$. The lower quartile does not change significantly.

\subsection{Manual intervention: The example of a super-group at $z = 0.22$} \label{seq:manual_intervention}

After applying our group-finding procedure to the real 10k sample, we encountered a huge structure which resisted to yield even roughly consistent results between FOF and VDM. We mention this particular case, because it illustrates some of the weaknesses of our adopted group-finder, and finally required a special treatment. It was already mentioned by Finoguenov et al. (2007) and is probably an example of a ``super-group'', where several smaller groups are just about to merge (Smol{\v c}i{\'c} et al. 2007). The redshift of the system is about $z \sim 0.22$. An example of such a super-group in another field at $z = 0.37$ is given by Gonzalez et al. (2005), Kautsch et al. (2008).

\begin{figure*}
		\epsscale{0.9}\plotone{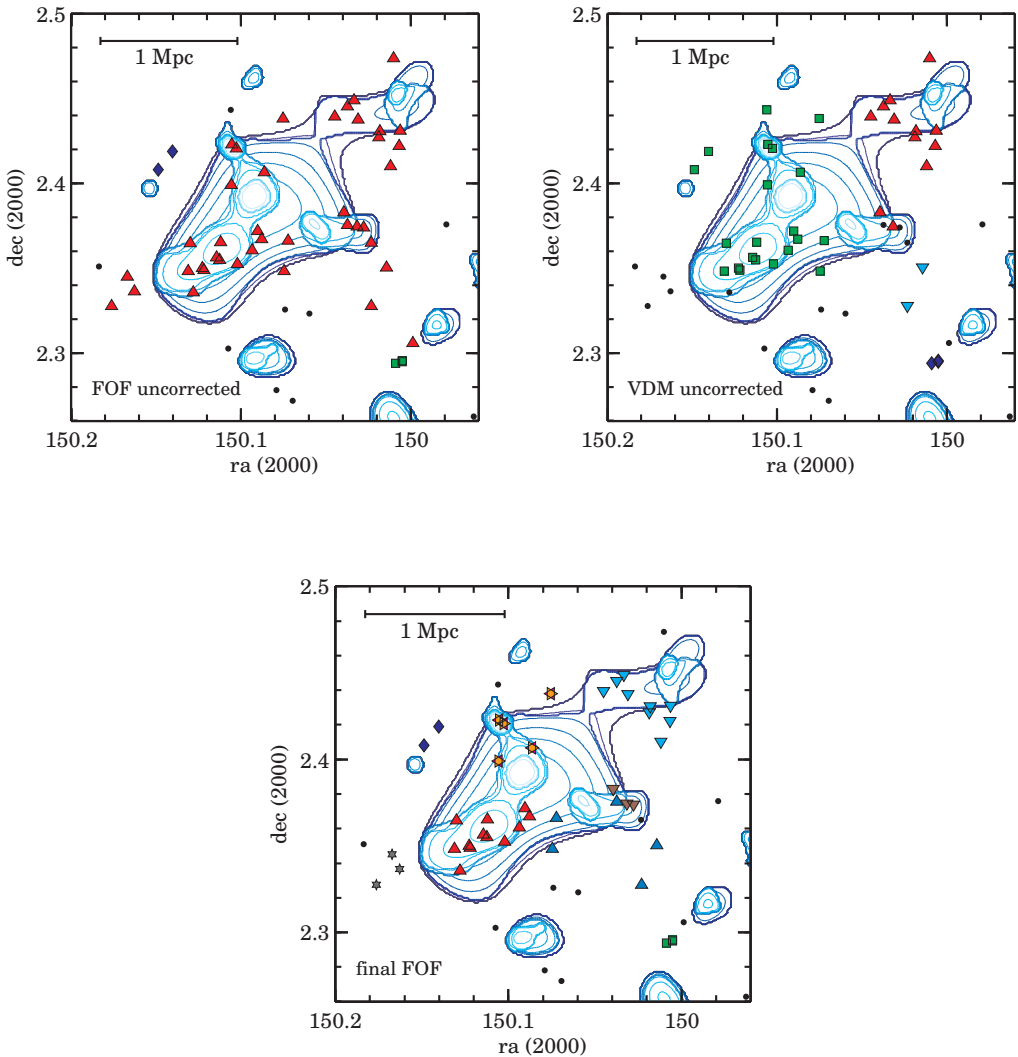}
		\figcaption{Group $\# 8$ in the uncorrected FOF group catalogue has been manually split up into several groups ($\# 8$, $\# 795- \# 799$), because the FOF as well as the VDM method failed (see the text). Upper left panel shows the initial FOF groups, the upper right panel the initial VDM groups, and the lower panel the FOF groups after manual intervention. Black points denote field galaxies, and the other symbols (squares, triangles, etc.) are group galaxies, whereas each group has its own symbol and color. The blue contours exhibit the X-ray emission of the super-group as observed with XMM-Newton (Finoguenov et al. 2007, 2009 in preparation).\label{fig:supercluster}}
\end{figure*}

The projected galaxy distribution of this structure is exhibited in Figure~\ref{fig:supercluster}.
The upper left panel shows the group assignment of the multi-run FOF catalogue, and the upper left panel the group assignment of the multi-run VDM. Each group is denoted by a symbol (e.g. square, triangle) of a particular color, and field galaxies by black points. This example of the super-group gives us some interesting insights concerning the group-finding procedure.

This extended structure exhibits the main potential problems of both the FOF and VDM algorithms. The FOF algorithm connected practically all the galaxies in this super-group, without distinguishing between different sub-groups. This behaviour is well known for FOF, and it happens in particularly dense regions such as this. The problem is that any single galaxy between two of these sub-clusters will act as a bridge for the FOF algorithm to connect the two clusters. The VDM is more successful in distinguishing different sub-structures, but nevertheless fails to do a perfect job. A casual glance suggests that the ``green square'' VDM cluster in fact consists of two independent sub-groups (consistent with the X-ray contours). Furthermore, the ``red triangle'' VDM group exhibits two outliers to the South which almost certainly do not belong to this group. The occurrence of such outliers is not uncommon in VDM groups. It is related to the fact that in the VDM groupfinder every second order Delaunay-neighbour in the second cylinder is accepted as group member and that the second cylinder is usually much bigger than the third cylinder (see \S~\ref{seq:VDM}).

Since we have accepted the FOF catalogue as the basis catalogue, and since this super-group is the only case where FOF is in such obvious disagreement with VDM, we decided to just correct this single structure manually after visual inspection. The final result is shown in the lower panel of Figure \ref{fig:supercluster}. The manually created groups by this intervention are group $\# 8$ and $\# 795-\# 799$. Although the manual assignment of galaxies to sub-groups is somewhat arbitrary, looking in redshift space (and not only at the projected galaxy distribution) the eye recognizes quite well different sub-structures.  This example emphasizes how important precise redshift information is. Even if it had been possible to recognize this huge overdensity with precise photo-z, it would not be possible to disentangle the more subtle sub-structure reliably. This is illustrated best by the ``blue upward triangle'' and the ``brown downward triangle'' groups overlapping in projection (see lower panel of Figure~\ref{fig:supercluster}).

\section{Comparison with the mocks and 2dfGRS}

In this section, we will compare our group catalogue with the mocks. Since zCOSMOS was designed to have a similar survey design as the 2 Degree Field Galaxy Redshift survey (2dfGRS, Colless et al. 2001)(see Lilly et al. 2007), albeit at higher redshifts, we also want to compare our group catalogue to the 2PIGG group catalogue (Eke et al. 2004) of 2dfGRS to have a reference point in the local universe. The 2PIGG catalogue is particularly appropriate for comparison since it was obtained by essentially the same FOF algorithm we have adopted.

\subsection{Number of groups as a function of $N$} \label{seq:n_of_N}

\begin{figure}
		\epsscale{1}\plotone{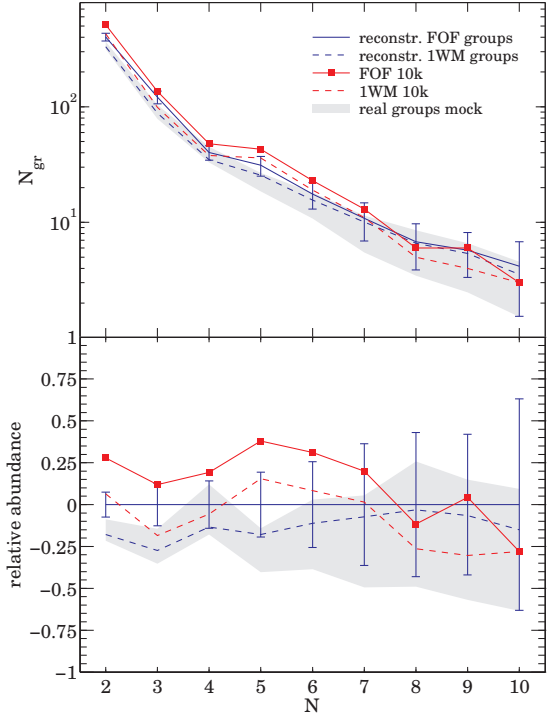}
		\figcaption{Number of groups $N_{\rm gr}$ as a function of observed richness $N$. The upper panel shows the absolute number of groups, whereas the red lines correspond to the 10k groups, and the blue lines to the reconstructed groups in the mocks. Solid lines correspond to FOF groups, and dashed lines to the 1WM groups. The errorbars correspond to the real scatter among the 24 mocks. The gray shaded area exhibits the cosmic variance of real groups in the mocks among the 24 mocks. The lower panel shows the abundance of groups relative to the reconstructed FOF groups, whereas all symbols are the same as in the upper panel.\label{fig:n_of_N}}
\end{figure}

The most straightforward way to compare the real data with the mocks is to compare the number of 10k groups with the number of reconstructed groups in the mocks as a function of observed richness $N$. This is shown in Figure~\ref{fig:n_of_N}.
The number of 10k groups is in good agreement with the number of reconstructed groups in the mocks. It seems that for $N \lesssim 6$ there is a slight excess of 10k groups relative to the reconstructed groups. However this excess is not significant.

More significant is the fact that for $N \geq 20$ there are significantly more groups in the mocks than in the real data. While in the 10k sample there is one group in this range (group $\# 2$ with $N = 20$)\footnote{Even if we regarded the super-group in \S~\ref{seq:manual_intervention} as a single group with $N>20$, the 10k group catalogue would still contain far fewer big groups than in most of the mocks.}, the median number of reconstructed groups with $N \geq 20$ in a 10k mock is 5 with upper and lower quartiles of 6 and 3, respectively. These groups are distributed in the redshift range $0.1 \lesssim z \lesssim 0.7$.  Even groups with $N > 50$ should not be exceptional --- on average there are $\sim 0.5$ of these per mock.

These huge groups are not an artefact of our reconstructed groups, but are also present in the real group catalogue. More than $80\%$ of the reconstructed groups with $N \geq 20$ exhibit a two-way-match to a real group, as one would expect from Figure~\ref{fig:statistics}, and their mean GRP$_1$ is about 0.9. So we conclude that the lack of big groups in the 10k sample is probably real and not due to some problem with the groupfinder.

\subsection{Fraction of galaxies in groups}\label{sec:fraction_of_galaxies_in_groups}

\begin{figure}
		\epsscale{1}\plotone{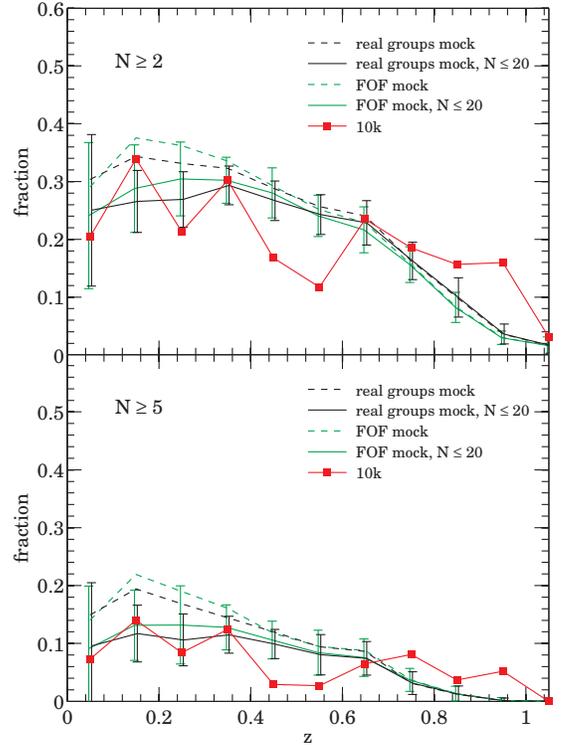}
		\figcaption{Fraction of galaxies in groups in the central region. The red solid line shows the fraction of galaxies in groups for the 10k catalogue. The black dashed line shows the mean fraction of galaxies in real groups in the mocks and the green dashed line the mean fraction of galaxies in reconstructed groups. If only galaxies in groups with $N\leq 20$ are considered as group galaxies, the results are the solid black and green line, respectively. The errorbars show the scatter among the 24 mocks.\label{fig:fraction_in_groups}}
\end{figure}

A quantity that is closely related to the number of groups in a catalogue is the fraction of galaxies in the sample that are placed in groups.  This fraction is shown as a function of redshift in Figure~\ref{fig:fraction_in_groups} for $N \geq 2$ and $N \geq 5$ in the central region.
If the galaxies in the mocks associated to groups with $N \geq 20$ are not considered as group galaxies (solid blue and black line) --- since we have shown that there are far too many of these groups in the mocks --- the overall behaviour of the fractions in the 10k sample (red line) match quite well those of the reconstructed groups or real groups, at least in the redshift range $0.1 \lesssim z \lesssim 0.5$. If the galaxies in groups with $N \geq 20$ are treated as normal group galaxies (dashed blue and black line), the fraction of group galaxies in the mocks is too high, as expected.

Noticeable is the lack of group galaxies in the 10k sample at redshift $z \sim 0.55$ and the excess at redshift $z \sim 0.9$, especially for $N \geq 2$. The lack of group galaxies at $z \sim 0.55$ coincides with a big underdensity in the 10k galaxy sample and is clearly visible in Figure~\ref{fig:n_of_z}. The origin of the excess of group galaxies at high redshift is less clear. While there are single mocks with a deviation at $z \sim 0.55$ similar to the lack of group galaxies in the 10k sample, none of the mocks approach the excess of group galaxies at redshift $z \sim 0.9$.  These underdense and over dense regions are clearly seen in the overall galaxy density field constructed by Kova\v{c} et al. (2009a).

\begin{figure}
		\epsscale{1}\plotone{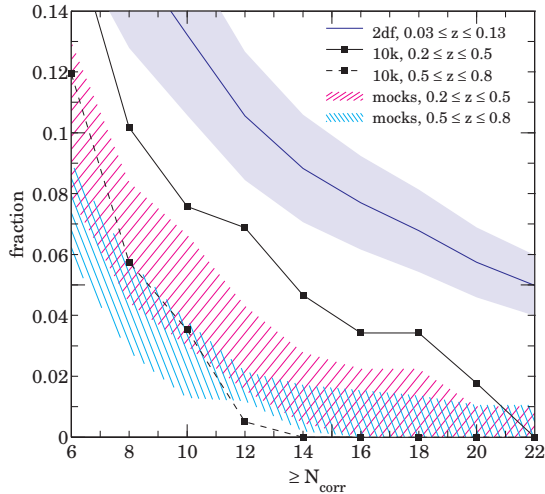}
		\figcaption{Fraction of galaxies in groups with equal or more than $N_{\rm corr}(-20)$ members in volume limited samples. The solid black line shows the fraction of 10k group galaxies in the redshift bin $0.2 \leq z \leq 0.5$, and the dashed black line the fraction in the redshift bin $0.5 \leq z \leq 0.8$, each time in the central region. The magenta and the cyan hatched regions display the regions enclosed by the upper and lower quartiles of the corresponding fractions in the mocks for the low and high redshift bin, respectively. The blue line displays the fraction for 2dfGRS-2PIGG groups in the redshift range $0.03 \leq z \leq 0.13$, and the blue shaded region is the uncertainty given by $\pm 0.2$ error in the absolute magnitude estimation of the 2dfGRS galaxies.\label{fig:n_eff_function}}
\end{figure}

The decline of group galaxies with redshift has two reasons: First of all, there are fewer groups at high redshift, as expected from the hierarchical growth of structure at these scales, and second, the fraction of detectable groups decreases with redshift, since the galaxy density decreases and so it becomes more and more improbable to observe two galaxies residing in the same DM halo. The second of these has been already discussed in \S~\ref{sec:detectability} (see Figure~\ref{fig:halo_incompleteness}), so we focus here mainly on the first reason which is demonstrated in Figure~\ref{fig:n_eff_function}.
Here the fraction of galaxies in groups is shown as a function of $N_{\rm corr}(-20)$, and all galaxies samples are chosen to be volume limited in respect to $M_{\rm b,lim}(z)$. The solid black line exhibits the fraction of 10k group galaxies in the redshift bin $0.2 \leq z \leq 0.5$, and the dashed black line the fraction in the redshift bin $0.5 \leq z \leq 0.8$, each time in the central region. The magenta and the cyan hatched regions show the regions enclosed by the upper and lower quartiles of the corresponding fractions in the mocks for the low and high redshift bin, respectively. The fractions in the mock are rather lower than in the 10k sample, especially at low redshift, although there are single mocks which have fractions as high as that in the 10k sample and higher. The blue line is the fraction of galaxies in the 2dfGRS-2PGIGG groups (Eke et al. 2004) in the redshift range $0.03 \leq z \leq 0.13$.

Note that the plotted lines are relatively sensitive to the absolute magnitudes used to estimated $N_{\rm corr}$. If there are slight systematics in the estimation of the absolute magnitudes, the lines will be slightly too low or too high. The absolute magnitudes for the 2dfGRS galaxies were estimated using the $k$-correction formula provided by Norberg et al. (2002). Furthermore, in order to adjust the 2dfGRS selection effects and completeness (Colless et al. 2001; Cross et al. 2004) as much as possible to those of zCOSMOS, some 2dfGRS galaxies were removed from the sample in a probabilistic way, and $N_{\rm corr}$ for 2PIGG groups were estimated in the same way as for 10k groups. The blue shaded region in Figure~\ref{fig:n_eff_function} exhibits the uncertainty for 2PIGG line owing for some possible systematics with the absolute magnitude estimation corresponding to $\pm 0.2$ mag. The effect from cosmic variance between NGP and SGP is much smaller than this.

Figure~\ref{fig:n_eff_function} shows clearly a decline of the fraction of group galaxies with redshift. Since $N_{\rm corr}$ correlates fairly well with mass (see Figure~\ref{fig:n_corr_mass}), this decline can straightforwardly be interpreted in terms of growth of structure as expected in a hierarchical structure formation scenario (Voit 2005). In fact, using $M_{b, \rm lim} = -20$ as absolute magnitude threshold to estimate $N_{\rm corr}$ our sample is volume limited up to $z \sim 0.9$, and we have checked with the mocks that we do not loose groups due to detectability problems in this sample.

\section{Summary}

The aim of this paper was to create a group catalogue out of the spectroscopic zCOSMOS 10k sample, to enable investigations of the galaxy population in groups over the redshift range $0.1 \lesssim z \lesssim 1$. The basic group-finding method was to use a FOF and a VDM groupfinder to identify galaxy overdensities in redshift space without regard to individual galaxy properties, using the precise $\sigma \sim 100$ km s$^{-1}$ velocities available from zCOSMOS.

The performance of both FOF and VDM groupfinders was extensively tested using realistic mock spectroscopic samples generated from the COSMOS mock lightcones (Kitzbichler \& White 2007), which reproduce the complex selection function of the actual spectroscopic survey. During the extensive testing and comparing of these groupfinders, we have developed a new method which progressively optimizes the group-finding parameters for smaller and smaller groups as the catalogue is generated from the richest groups down to the poorest. This is found to optimize the group catalogue fidelity, in terms of completeness and purity, over a broad range of richnesses $N$. Using this new approach, we achieve an impressively high fidelity of our group catalogue compared with others in the literature at these redshifts.  The standard FOF algorithm yields a group catalogue with overall better statistics compared with those of the VDM algorithm, and we have chosen this for basic group catalogue. However, the purity of the group sample is significantly enhanced, at modest cost in completeness, if we also take the intersection of this FOF main catalogue with an independent VDM group catalogue --- producing the so called ``one-way-matched'' (1WM) sub-catalogue.  

With the aid of our mocks, we have a very good idea of the statistical properties of the group sample. We find that for FOF groups with $N \geq 5$ the completeness is $85 \%$ and the purity $78 \%$. For the 1WM catalogue, the purity rises to $82 \%$, while the completeness drops only to $81 \%$. For poorer groups with $N<5$ the statistics of purity and completeness are not substantially worse. These fidelity statistics are fairly stable over the whole redshift range. As would be expected, the completeness and ''interloper fraction'' statistics for group members are enhanced in the centers of groups. Furthermore, we find that, while the basic FOF catalogue slightly overproduces the number of groups with $N \lesssim 5$, the 1WM sub-catalogue reproduces almost perfectly the number of real groups down to $N = 2$.

The actual zCOSMOS 10k FOF group catalogue contains 102 groups with $N \geq 5$ and 23 groups with $N \geq 8$. Going down to $N = 2$ yields a total of 800 groups. Groups with $N \geq 5$ have been assigned a velocity dispersion $\hat \sigma$ and a mock calibrated dynamical mass $\hat M$ whose uncertainties are understood quite well. While for $N<5$ we could still assign a meaningful mass to the groups, a reasonable estimate of velocity dispersion is not possible. The fraction of 10k galaxies in groups is about $25 \%$ at low redshift and decreases toward $\sim 15 \%$ at $z \sim 0.8$.

Comparing the 10k group catalogue to the mocks yields fairly consistent results. The main discrepancies are that (1) there are many more groups with $N \geq 20$ in the mocks compared to the 10k sample, and (2) the fraction of 10k galaxies in groups is significantly higher at $z \sim 0.9$ than in the mocks. We find that the fraction of galaxies in groups, for groups with a given corrected richness $N_{\rm corr}(-20)$, decreases from redshift 0.1 to redshift 0.8. This can be interpreted in terms of growth of structure as expected in a hierarchical structure formation scenario.

The properties of these groups are explored in a number of companion papers. The environments of these groups in terms of the larger scale galaxy-density-field in which they are embedded is given in Kova\v{c} et al. (2009a). The evolution of the galaxy population in these groups is explored in A. Iovino et al. (2009, in preparation) and K. Kova\v{c} et al. (2009b, in preparation), in terms of galaxy colors and galaxy morphologies respectively, and in Silverman et al. (2009) in the context of active galactic nuclei. In future studies, we will investigate the X-ray properties of our groups, and perform a weak lensing and a galaxy-group cross-correlation analysis.

\section{Acknowledgements}

We thank Peder Norberg and Oliver Hahn for useful discussions.  This research was supported by the Swiss National Science Foundation, and it is based on observations undertaken at the European Southern Observatory (ESO) Very Large Telescope (VLT) under the Large Program 175.A-0839.

\end{document}